\documentclass[12pt,preprint]{aastex}

\usepackage{epsfig}

\shorttitle{Planetary Nebulae in GLIMPSE~II}
\shortauthors{Zhang \& Kwok}

\begin{document}

\title{Planetary Nebulae Detected in the {\it Spitzer Space Telescope} GLIMPSE~II
Legacy Survey\footnote{ Figures~2--43 are available on the ApJ website or
on request from the authors.}}

\author{Yong Zhang \& Sun Kwok}
\affil{Department of Physics, University of Hong Kong, Pokfulam Road, Hong Kong}
\email{zhangy96@hku.hk; sunkwok@hku.hk}

\begin{abstract}

We report the result of a search for the infrared counterparts of 37 planetary nebulae (PNs) and PN candidates 
in the  {\it Spitzer} Galactic  Legacy Infrared Mid-Plane Survey Extraordinaire II (GLIMPSE~II) survey.  The photometry and images of these PNs at 3.6, 4.5, 5.8, 8.0, and 24\,$\mu$m, taken through the Infrared Array Camera (IRAC) and the Multiband Imaging Photometer for Spitzer (MIPS), are presented.
Most of these nebulae are  very red and compact in the IRAC bands, and are found to be bright and extended in the 24\,$\mu$m band.
The infrared  morphology of these objects are compared with H$\alpha$ images of the Macquarie-AAO-Strasbourg (MASH) and MASH~II PNs. The implications for morphological difference in different wavelengths are discussed.  
The IRAC data allow us to differentiate between PNs and \ion{H}{2} regions and be able to reject non-PNs from the optical catalogue (e.g., PNG\,352.1$-00.0$).
Spectral energy distributions (SEDs) are constructed by combing the IRAC and MIPS data with existing near-, mid-, and far-IR photometry measurements.  The anomalous colors of some objects allow us to infer the presence of aromatic emission bands.
These multi-wavelength data provide useful insights into the nature of different nebular components contributing to the infrared emission of PNs. 

\end{abstract}
\keywords{infrared: ISM --- planetary nebulae: general ---  stars: AGB and 
post-AGB}

\section{INTRODUCTION}

The traditional view of planetary nebulae (PNs) is that it is a shell of gas ionized by a hot central star and exhibits an emission-line spectrum.  The theoretical understanding that PNs descend from asymptotic giant branch (AGB) stars and should contain remnants of the dust envelope led to the realization that PNs should also be bright infrared objects \citep{kwok82}.  This was confirmed after the {\it Infrared Astronomical Satellite} ({\it IRAS}) all-sky survey found that over 1000 PNs show far infrared emission \citep{pot84, zk91}.  The typical color temperature of the dust component in PNs is between 100 and 200 K, warmer than the dust component in H~{\sc ii} regions but cooler than the circumstellar dust envelopes of AGB stars.  This is consistent with the fact that the dust envelopes in PNs represent the dispersing remnants of the AGB envelopes \citep{kwok90}.  Infrared emission is now recognized as a defining observational properties of PNs.

However, the location of the infrared emitting dust component remains poorly known.  Only recently has the angular resolution of mid-infrared cameras  on large ground-based telescopes with adaptive optics become high enough to image the  distribution of dust in PNs.  Space observations therefore offer a better opportunity to survey a large number of PNs.  The  Infrared Array Camera (IRAC)  on the  {\it Spitzer Space Telescope} has an angular resolution $<2$ arcsec \citep{fazio04} and its capability in imaging the dust component of PNs has been  clearly demonstrated by \citet{hora04}.  Since the peak of the dust emission from PNs occurs at wavelengths $>20$ $\mu$m, the Multiband Imaging Photometer for Spitzer \citep[MIPS;][]{rieke04}, although offering less angular resolution than IRAC, would provide useful information on the distribution of the main cold dust component.  Based  on the MIPS observations at 24 and 70\,$\mu$m,
\citet{su07} found possible evidence for the presence of a debris disk around the central star of the Helix Nebula.  Other examples of PNs studied by {\it Spitzer} includes NGC\,2346 \citep{su04}, NGC\,650 \citep{ueta06}, and a sample of PNs in the Large Magellanic Cloud \citep[LMC;][]{hora08}.

One of the advantages of IR observations is that they are hardly affected by dust reddening and are therefore ideal to study PNs heavily obscured by interstellar dust.  From the IRAC observations, \citet{cohen05a} identified
an extremely red PN G313.3+00.3 which is optically invisible.  This raises the possibility that there are many obscured
PNs in the Galaxy, for which {\it Spitzer} IR observations could explore.  A high-spatial-resolution and high-sensitivity H$\alpha$ survey of the southern Galactic plane has been undertaken using the Anglo-Australiam Observatory UK Schmidt Telescope (AAO/UKST) to search for very faint PNs \citep{par99}.  This has led to the discovery of over 1000 new PNs and PN candidates (which we refer to hereafter simply as PNs).  The results of this survey was published as the Macquarie/AAO/Strasbourg H$\alpha$ Planetary Galactic Catalogue (MASH) \citep{parker06} and the MASH~II supplement \citep{miszalski08}.  These new PNs are typically more obscured than known Galactic PNs and
warrant further investigation at IR wavelengths.

The {\it Spitzer} GLIMPSE (Galactic Legacy Infrared Mid-Plane Survey Extraordinaire; see Benjamin et al. 2003 and Churchwell et al. 2009) survey provides IRAC images with high resolution and sensitivity at wavelengths from 3.6 to 
8.0\,$\mu$m, allowing us to perform high-quality infrared imaging and photometry of PNs. The first-release GLIMPSE~I data cover a Galactic region of $\left|l\right|=10^\circ$--65$^\circ$ and $\left|b\right|\le1^\circ$ (total area $\sim220$ square degrees). Comprehensive GLIMPSE~I-based studies of PNs have been presented by some authors \citep[e.g.][]{cohen07b,phillips08a, phillips08b,ramos08}.
\citet[][hereafter Paper~I]{kwok08} imaged 30 PNs\footnote{Note
that some objects in Paper~I are actually compact \ion{H}{2} regions, but are
referred to as `PNs' for convenience of denotation (their non-PN nature has
been discussed in text).} 
 detected in the GLIMPSE~I survey and constructed
IR spectral energy distributions (SEDs) which clearly show
a distinction between the photospheric, nebular bound-free, and dust 
emission components.

In this paper, we extend the work of Paper~I to the GLIMPSE~II survey
area that covers the Galactic bulge regions. We present the photometry
and images of 37 PNs detected in the GLIMPSE~II survey. 
These results are compared with those of the GLIMPSE~I PNs.
To better explore nebular dust emission,
we also take into account the 24\,$\mu$m data
from the {\it Spitzer} Legacy program MIPS Inner Galactic Plane
Survey (MIPSGAL). This paper is organized as follows:
in Section~2 we briefly introduce the GLIMPSE~II observations
and the {\it Spitzer} data used in this study;
in Section~3 we describe  the data processing;
in Section~4 we present the Mid-IR images and
our measurement results of fluxes, sizes, and SEDs
for individual PNs; the astrophysical implications 
of our results are discussed in Section~5; and 
Section~6 summarizes the main conclusions.

\section{OBSERVATION}

The data used in this paper were taken
from the  GLIMPSE~II Version 2.0 Data Release\footnote{see 
http://www.astro.wisc.edu/glimpse/glimpsedata.html.}.
The GLIMPSE~II survey covers longitudes
within 10$^\circ$ of the center regions of the Galaxy
($\left|b\right|\le1^\circ$ for $\left|l\right|=10^\circ$--5$^\circ$, 
$\left|b\right|\le1.5^\circ$ for $\left|l\right|=5^\circ$--2$^\circ$, 
and $\left|b\right|\le2^\circ$ for $\left|l\right|=2^\circ$--0$^\circ$;
total area $\sim54$ square degrees) 
at 3.6, 4.5, 5.8, and 8.0\,$\mu$m with resolutions from
1.6$''$--1.8$''$, but excludes an area of $\sim3$ square degrees
at the Galactic center observed by the PID 3677 GO program (which
however has been included in the GLIMPSE~II v2.0 mosaics). 
The survey regions are roughly shown in Figure~\ref{lb}. The observations 
were made with three 1.2 second exposures at each position 
during the periods between September 2005 and April 2006.
The procedures of data reduction are similar to those described
in Paper~I. 
The GLIMPSE~II data products include mosaic images, a
highly reliable Point Source Catalog (GLMIIC), and
a more complete Point Source Archive (GLMIIA).
Further details of the GLIMPSE~II data can be found in \citet{churchwell07}
and \citet{mea08}.


We also made use of the data from the MIPSGAL survey\footnote{see
http://mipsgal.ipac.caltech.edu/.}.  The MIPSGAL observations
were carried out at several epochs between 2005--2006
utilizing  the MIPS instrument aboard {\it Spitzer},
aiming at providing a longer-wavelength complement to the 
GLIMPSE survey. An area of 278 square degrees in the inner Galactic plane,
completely overlapping the GLIMPSE~II observations,
was imaged at the 24 and 70\,$\mu$m bands.  In this study, we only used the 
publicly available 24\,$\mu$m images of the same region as GLIMPSE~II. These
images have a basic sensitivity of $\sim110$\,$\mu$Jy and a field of view 
(FOV) of $\sim5.4'\times5.4'$.  The detector for MIPS24 is $128\times128$ pixel 
Si:As arrays with a bandwidth of 4.7\,$\mu$m.
The 24\,$\mu$m mosaics have a saturation limit of
1700\,MJy/sr for extended source and a resolution of
$6''$.  The data calibration was performed by both
 the {\it Spitzer} Science Center (SSC) and the MIPSGAL team.
A good summary of the current status of the MIPS24 data sets
can be found in \citet{cohen09}.
The data reported in this paper were taken from
the MIPSGAL v3.0 Data Delivery \citep{carey08}, which provides
the 24\,$\mu$m mosaics ($1.1^\circ\times1.1^\circ$ in size) for all the 
survey regions.

\section{DATA PROCESSING}

We compared the GLIMPSE~II data sets with the Edinburgh/AAO/Strasbourg Catalogue of Galactic Planetary Nebulae
\citep[version 1.0;][]{parker01} and the MASH/MASH~II catalogue (hereafter referred to as MASH PNs).
These optical catalogues include the new PNs identified in the AAO/Strasbourg H$\alpha$ survey of the Milky Way
and a small percentage of previous known Galactic PNs.  We first selected a subsample of 71 MASH PNs lying in the GLIMPSE~II survey region, among which 10 were taken from the MASH~II catalogue. We then undertook careful visual examination of the GLIMPSE~II images in all the four IRAC bands using the DS9 software package\footnote{DS9 is
developed by Smithsonian Astrophysical Observatory.}.  If a PN is visible at at least one of the bands, we extracted a
portion of the four-band images centered on the coordinates of this PN and measured its integrated fluxes.


The total fluxes of these PNs were obtained using the same method as in Paper~I. 
We first deduced the total on- and off-source fluxes using two apertures which
have an identical size and were respectively centered on the PNs 
and in a position near the sources.
Then the sum of all the fluxes of sources in the point-source 
archive within each aperture was obtained and subtracted from
the on- and off-source fluxes respectively. The PN fluxes were finally
determined from the on-source net fluxes subtracted off-source net fluxes.
We have taken care to avoid over- and under-subtraction of the fluxes of 
point sources. For saturated sources, we only estimated lower limits of the fluxes.  Since most of the GLIMPSE~II PNs have well-defined boundaries, we also measured the major and minor diameters of each PN. For a few sources which have diffuse structures, the sizes of the bright parts were roughly estimated.

The same procedure is also applied to the MIPS24 data. There are several MASH PNs which have prominent 24\,$\mu$m emission but have no IRAC counterparts. We do not include these PNs in this analysis.  Given the fact that the MIPS24 images have a resolution about three times lower than the IRAC ones, in some cases field stars cannot be
resolved from the PNs in the  MIPS24 images, and thus the flux measurements should be treated 
with some caution. The 24\,$\mu$m mosaics show some artificial structures around bright sources. We have attempted to minimize the effects of artifacts on the measurements of fluxes and sizes. Some PNs show faint extend halos which are likely to be real.  We did not take into account these faint halos in the flux estimation since the fluxes of these halos are negligible compared with those of the main nebular regions.

\section{RESULTS}

We find that 37 MASH PNs (including five from the MASH~II catalogue) have visible IRAC counterparts, among which nine are previously known Galactic PNs (Table\,\ref{name}).  The detection rate of the GLIMPSE~II PNs is 52$\%$, about three times higher than that of the GLIMPSE~I PNs (18$\%$; Paper~I). This is partly ascribed to the fact that the interstellar extinction toward the Galactic center is heavier and thus  the optical observations only revealed
those relatively bright PNs within the GLIMPSE~II field.  Most of the GLIMPSE~II PNs appear to be compact and spherical, probably suggesting that they are generally more distant than GLIMPSE~I PNs although we cannot completely rule out the possibility that some GLIMPSE~II PNs are young and intrinsically compact.  All the GLIMPSE~II PNs are strong 24\,$\mu$m sources, some of which are even saturated.

Figure.~\ref{lb} gives the approximate survey coverage of the GLIMPSE~I/II  projects
overlaid with the positions of MASH PNs. 
A total of 12 GLIMPSE~II PNs are located within the regions of 
$\left|b\right|\le1^\circ$. This corresponds to a surface density of
$N_{\rm PN}=3$\,PNs per 10\,deg$^2$, a factor of six lower than that of GLIMPSE~II 
PNs in the regions of $\left|b\right|>1^\circ$, where $N_{\rm PN}=18$\,PNs per 10\,deg$^2$.
This is due to the strong IR background emission which hampers
the detections of those PNs close to the Galactic plane, as one
can clearly see in \citet{mea07} (their Figure.~1).
This suggests that IRAC counterparts are more easily found outside the galactic plane.

In Figure.~\ref{lb} we also plot the coverage of the GLIMPSE~3D survey (PI: Benjamin; PID$=30507$), which
images 112 square degrees in a series of regions with higher Galactic latitudes in all the four IRAC bands aiming at investigating the vertical stellar and interstellar structure of the inner Galaxy. We suppose that more PNs can be discovered in the GLIMPSE~3D field and it is interesting 
to compare the properties of these PNs located at higher Galactic 
latitudes with those of the GLIMPSE~I/II PNs. Our work on the PNs detected in the GLIMPSE~3D survey is in progress and will be the subject of a forthcoming paper.

The IRAC images of the 37 PNs are displayed in Figure~\ref{irac}--\ref{irac37}, where we also overplot the contours of
H$\alpha$ emission for comparison. 
The 24\,$\mu$m images of these objects are displayed in Figure~\ref{mips}--\ref{mipsend}. 
In order to better show the features at different bands, we have adjusted the imaging depths as well as contrasts
to create these color images.
It is evident from these figures that most of these PNs appear to be redder and more extended than the surrounding 
field stars.  The mid-IR (MIR) photometry results are presented in Table.~\ref{flux}.
The first column lists the source names which  are designed as ``PNG $lb$'' following the IAU convention. 
Column~2 and 3 give the source right ascensions and declinations (J2000),
which are mostly taken from the MASH catalogues.  In a few rare cases the MASH catalogues give incorrect
coordinates (e.g. PNG $000.9+01.8$) the corrected values are given in this table.  Column~4--8 give the spatially integrated fluxes in the 3.6--24\,$\mu$m bands. Due to the very bright background emission that tends to overwhelm the near-IR emission from the PNs, most of the PNs near the Galactic center were detected by IRAC only in the 8.0\,$\mu$m band. 
The measured source sizes are presented in Column~9--13.  For comparison, column~14 gives the H$\alpha$ sizes taken from
the MASH catalogues. For the PNs that are detected in all the four IRAC bands, we calculated the 3.6--8.0\,$\mu$m SED slopes
using $\alpha_{\rm IRAC}=d\log(\lambda F_\lambda)/d\log(\lambda)$.  The resultant $\alpha_{\rm IRAC}$ values are given in
column~15.

\subsection{Spectral Energy Distributions}

To better constrain the IR spectral properties of the GLIMPSE~II PNs, we have searched other data archives, including the Two Micron All Sky Survey (2MASS), Deep Near Infrared Survey of the Southern Sky (DENIS),
{\it Midcourse Space Experiment (MSX)}, and the {\it IRAS} Point Source Catalogue (PSC).  
Data from these catalogues together
cover the spectral region from 0.82  to 100\,$\mu$m. All the available images have been checked visually
to ensure that the correct counterpart is used. Table~\ref{other} tabulates  the magnitudes and fluxes of these measurements.   As these other surveys all have lower sensitivities than IRAC, in general only the brightest sources in the GLIMPSE~II sample are detected in the other surveys.  Since IRAC also offers better angular resolution, all the sources 
are unresolved in these archives, with the only exception of PNG $352.1-00.0$.
As PNG $352.1-00.0$ has a large angular size, the data given in these point source catalogues only represent the lower limits of fluxes.  We therefore  do not include this object in Table~\ref{other}.  Combining our IRAC and MIPS photometry  with these archive data, we have constructed SEDs for the PNs listed in Table~\ref{other}  and the results are displayed in Figure~\ref{sed}.

A glance at Figure~\ref{sed} shows that almost all the GLIMPSEII PNs are red objects with a rising spectrum between 1 and 20 $\mu$m.  Their SEDs are in general consistent with the presence of a cold dust component of temperature lower than 300\,K.
This confirms the {\it IRAS} observations that PNs are generally surrounded with cold dust shells \citep{zk91}.
Some sources have a separate emission component at short wavelengths
($<3$\,$\mu$m), which can be attributed to contributions from the photospheric, nebular bound-free, or hot dust emission components \citep[see,][for a detailed discussion of SEDs of PNs]{zk91}. 

The SED slope $\alpha_{\rm IRAC}$  can provide quantitative description of the SED shape.  We have determined $\alpha_{\rm IRAC}$ through least-squares fit to the four IRAC bands, and their values are listed in Table~\ref{irac}. The SED slopes for all the GLIMPSE~II PNs are $ \alpha_{\rm IRAC}>-1$, substantially larger than those found for stellar photospheres \citep{lada06,chavarria08}.  Based on their study of young stellar objects (YSOs), \citet{chavarria08}
introduced a classification scheme: Class~I for  $ \alpha_{\rm IRAC}>0$
(star plus infalling envelope), Class~II for $-2\ge\alpha_{\rm IRAC}\le0$
(star plus a disk), and Class~III for $ \alpha_{\rm IRAC}<-2$
(post-T Tauri stars or main-sequence photospheres). According to
this classification, we found that among the 20 PNs with
 $\alpha_{\rm IRAC}$ measurements 80 percent are
categorized as Class~I (star with envelope), and 20 percent as Class~II
(star with disk). \citet{fuente08} determined the $\alpha_{\rm IRAC}$ 
values for 19 ultracompact (UC) H~{\sc ii} regions, which are comparable with those for PNs.  Although further statistical work based on a larger PN sample needs to be performed to determine whether the classification of YSO can 
apply to PNs, we show here that $\alpha_{\rm IRAC}$ can serve as a useful indicator to investigate nebular properties. Nevertheless, it should be borne in mind that the slopes of PNs' SEDs can be significantly affected by various emission features (as discussed below) and it is difficult to reach solid conclusions without spectroscopic
studies.

Although the SEDs are useful to distinguish contributions from the photospheric, nebular, and dust components, the presence of various emission features complicates the IRAC/MIPS based studies of nebular IR properties.
These possible emission features include: the contributions by the 3.3, 6.2 and 7.7 $\mu$m aromatic infrared bands (AIB) to the  3.6, 5.8, and 8.0\,$\mu$m IRAC bands, molecular hydrogen vibrational-rotational bands to all the IRAC bands, recombination lines (Pf$\gamma$ 3.7\,$\mu$m, Br$\alpha$ 4.05\,$\mu$m, and Pf$\alpha$ 7.5\,$\mu$m) to the 3.6, 4.5, and 
8.0$\mu$m bands, and atomic collisionally excited lines to the 4.5--24\,$\mu$m bands
(see also Paper~I).  
It is impossible to accurately estimate the contributions of these features without spectroscopy.  The  {\it ISO} spectrum of PNG $359.3-00.9$ (Hb~5) is available \citep{pottasch07} and is plotted in Figure.~\ref{sed}. Figure.~\ref{resp} gives an expanded view of this spectrum overlaid with the spectral response curves for the IRAC and MIPS 24\,$\mu$m bands.  This figure shows that the contributions from these features to these bands  are 
significantly large to change the MIR colors.

\subsection{MIR Colors}

In Figure~\ref{color} we plot the $[3.6]-[4.5]$ versus  $[5.8]-[8.0]$ color-color diagram for the GLIMPSE~II PNs. For comparison, this figure also plots the colors of the GLIMPSE~I PNs reported in the literature \citep[Paper~I;][]{hora04,cohen07a,cohen07b,phillips08a,phillips08b}.  There is no obvious color difference between the GLIMPSE~I and the
GLIMPSE~II PNs in this diagram.  These PNs are very red and clustered around a position near $(2,1)$.  For the PNs with MIR emission dominated by contributions  from the central stars, the IRAC colors  are close to the $(0,0)$ point.
We have not attempted to correct for interstellar extinction towards these PNs  (which is supposed to be small at IRAC wavelengths) but show a reddening vector of $A_V=10$\,mag in this figure.  We note that the $[5.8]-[8.0]$ color index decreases with increasing reddening because extinction in the 8.0\,$\mu$m band  is primarily caused by a silicate absorption feature peaking at 9.7\,$\mu$m. Although one expects Galactic bulge PNs to suffer a higher degree of interstellar extinction, but this is not evident in Figure~\ref{color}.  The GLIMPSE~II PNs are indistinguishable from the GLIMPSE~I PNs in the color-color diagram. Furthermore, the data scattering direction in Figure~\ref{color} is roughly perpendicular to the reddening vector. We thus conclude that MIR extinction is negligible and does not affect the colors of the PNs in our sample. The IRAC colors of PNs mainly reflect the color temperatures and the relative contributions from different components including the photosphere, the ionized nebula, the dust envelop, and various emission features.  

We have calculated the simulated colors using blackbody models convolved with the spectral response functions,
and the results are plotted in Figure~\ref{color}.   Using the simulated blackbody colors as a reference, we can see that the $[3.6]-[4.5]$ colors of PNs suggest significantly higher temperatures than those by the $[5.8]-[8.0]$ colors. This is indicative of the substantial contribution of emission from 
AIB emissions, ionized gas, and/or photosphere to the 3.6\,$\mu$m band.
The relative contribution of the latter two have been illustrated
in \citet{zk91}.

In spite of the fact that PNs have different colors with stellar objects, the color-color diagram alone does not enable to completely distinguish PNs from other MIR sources.  From Figure~\ref{color} we can find that the colors for PNs partly overlap with those for Class~I YSOs (i.e. star with envelope) and the PNs with strong emission from central stars 
might be confused with Class~II YSOs (i.e. star with disk).  Moreover, some H~II regions are located within the same position as PNs in the color diagram. 
Therefore, despite the color-color diagram can serve as a tool to sample PN candidates, detailed investigations of morphologies and spectra are essentially required for accurate PN designation.

Figure~\ref{color2} shows the $[3.6]-[8.0]$ versus  $[8.0]-[24]$ color-color diagram. The $[3.6]-[8.0]$ colors are in the range from 1--5, and the $[8.0]-[24]$ colors are in the range from 3--7.  The IRAC-MIPS colors for these Galactic PNs
are similar to those found for LMC PNs by \citet[][see their Figure~5]{hora08}. For the same reason as in Figure~\ref{color}, the PN colors cannot be explained by single-temperature blackbodies.
Comparing Figure~\ref{color2} with the color-color diagram for the members of the IC~348 cluster presented by 
\citet[][see their Figure.~7]{lada06}, we find that the distribution of PNs in this diagram are clearly centered at a position redder than IC~348 cluster sources, even for the YSOs with thick disk. This suggests that the 24\,$\mu$m excess is a prominent characteristic of PNs.

In Fig.~\ref{colmag} we plot the IRAC color-magnitude diagrams ($[3.6]$ versus $[3.6]-[8.0]$ and $[8.0]$ versus $[3.6]-[8.0]$).  These diagrams show that the GLIMPSE~II PNs generally have lower integrated fluxes than the GLIMPSE~I PNs, although their colors are similar  to each other.  These diagrams can be compared with those for LMC PNs \citep[][Fiugre~6]{hora08}.  We find that the distribution of Galactic PNs in the color-magnitude diagrams perfectly follows those of LMC PNs in shape although the galactic PNs are nearer and hence brighter.

\subsection{Individual Objects}

PNG 000.0$-01.3$.--- This source was detected by IRAC in the 5.8 and 8.0\,$\mu$m bands only (Figure~{\ref{irac}}).  \citet{parker06} classified it as a `Likely' PN.   The object was also detected by the  NRAO VLA Sky Survey \citep[NVSS;][]{condon98}, which measured a flux density of $4.5\pm0.8$\,mJy at 1.4\,GHz.
In spite of its faintness,  its  H$\alpha$  and 8.0\,$\mu$m images are visually similar and both reveal an oval structure. 

PNG 000.1$-01.7$.--- This source was detected by IRAC in the 8.0\,$\mu$m 
band only (Figure~{\ref{irac2}}). Its PN status has been confirmed
by spectroscopy \citep[designated as JaSt85;][]{jacoby04}. \citet{steene01} 
observed this Galactic bulge PN in the radio continuum at 3 and 6\,cm and
estimated a distance of $\sim7.2$\,kpc. The NVSS  survey
gives a flux density of  $2.9\pm0.6$\,mJy at 1.4\,GHz
\citep{condon98}.  Given large [\ion{N}{2}]/H$\alpha$
flux ratio \citep{parker06}, it might be a low excitation PN.
It is extremely faint in the IRAC bands but fairly bright
in the MIPS24 band. Albeit with an ambiguous boundary, the size of 
8.0\,$\mu$m image is roughly consistent with that of H$\alpha$.

PNG 000.3$-01.6$.--- This source was detected by IRAC in the 8.0\,$\mu$m 
band only (Figure~{\ref{irac3}}).  Its PN status has been confirmed
by spectroscopy \citep[designated as JaSt86;][]{jacoby04}. Based the radio
observation, \citet{steene01} estimated a distance of 6.6\,kpc.
 The NVSS  survey gives a flux density of  $4.1\pm0.5$\,mJy at 1.4\,GHz
\citep{condon98}.
It is very compact in all the MIR and H$\alpha$ bands. 


PNG 000.5$+01.9$.--- This source was detected by IRAC in the 8.0\,$\mu$m 
band only (Figure~{\ref{irac4}}). The object was first discovered
by \citet{kohoutek94} (designated as K~6-7), and its PN status was 
subsequently confirmed by spectroscopy \citep[designated as JaSt17;][]{jacoby04}.
\citet{steene01} estimated the distance to be 7.1\,kpc. 
 The NVSS  survey gives a flux density of  $6.8\pm0.9$\,mJy at 1.4\,GHz
\citep{condon98}. The 8.0\,$\mu$m and H$\alpha$ images are visually similar
and show a compact and oval structure.

PNG 000.9$-01.0$.--- This source was detected by IRAC in the 8.0\,$\mu$m 
band only (Figure~{\ref{irac5}}). It is classified as a `True' PN
in the MASH~II catalogue. 
The MIR ( 8.0 and 24\,$\mu$m) and H$\alpha$ bands show a compact and round 
structure without a sharp boundary.

PNG 000.9$+01.8$.--- This source was detected by IRAC in the 8.0\,$\mu$m 
band only (Figure~{\ref{irac6}}). \citet{parker06}
classified it as a `True' PN. The optical image shows an oval
structure.  From the IRAC image, we note that
the object seems to be a core
surrounded by an ambiguous diffuse halo.

PNG 001.0$-01.9$.--- This source was detected by IRAC in the 8.0\,$\mu$m band only (Figure~{\ref{irac7}}). The object was first discovered by \citet{kohoutek02} (designated as K~6-35) and  classified as a `True' PN by \citet{parker06}.  The NVSS  survey gives a flux density of  $4.8\pm0.6$\,mJy at 1.4\,GHz \citep{condon98}. The H$\alpha$ image shows that it is a bright, compact, and slightly oval PN with faint
outer halo. Its 8\,$\mu$m counterpart is nearly overwhelmed by bright
background emission.

PNG 001.0$+01.9$.--- This source was only marginally detected by IRAC in the 
8.0\,$\mu$m band and was completely invisible in the other IRAC bands
(Figure~{\ref{irac8}}).  \citet{exter04} determined chemical abundances of this object with optical spectroscopy and classified it as a Type~I PN.
The NVSS survey gives a flux density of $20.6\pm1.3$\,mJy at 1.4\,GHz \citep{condon98}.  \citet{zhang95} deduced the statistical distance to be 3.46\,kpc. In their catalogue of narrow band images of PNs,
\citet{schwarz92} presented H$\alpha$ and [\ion{O}{3}] images of this PN, which exhibit an oval appearance with a size of 74$''$ and of 43$''$, respectively.
The 8.0\,$\mu$m image is too faint to obtain a reliable flux.
This PN is different from others in  this  sample of PNs in that its emission is significantly more compact at 24\,$\mu$m than at H$\alpha$.

PNG 001.6$-01.1$.---  This source was detected by IRAC in the 8.0\,$\mu$m
band only (Figure~{\ref{irac9}}). Its PN status has been confirmed
by spectroscopy \citep[designated as JaSt97;][]{jacoby04}.
\citet{steene01} estimated the distance to be 5.9\,kpc. 
The 8.0\,$\mu$m and H$\alpha$ images have a similar size and
both exhibit a compact oval  appearance.

PNG 002.0$+00.7$.---  This source was detected by IRAC in  all the four
bands (Figure~{\ref{irac10}}). The NVSS survey gives a flux density of 
$4.1\pm0.5$\,mJy at 1.4\,GHz \citep{condon98}. 
It is classified as a `True' PN in the MASH~II catalogue. The PN appears to be a point source in all the
bands. Its IRAC color is only slightly redder than stars. The SED indicates the presence of cold dust with $T<300$\,K (Figure~\ref{sed}).

PNG 002.1$-00.9$.---  This source was detected by IRAC in  all the four bands (Figure~{\ref{irac11}}). \citet{parker06} classified it as a `Likely' PN. It was also independently found by \citet{kohoutek02}.
This PN is very bright in all the bands, and exhibits a compact and round appearance. The SED indicates a cold dust component with $T<300$\,K (Figure~\ref{sed}).

PNG 002.1$-01.1$.---   This source was detected by IRAC in the 5.8 and 8.0\,$\mu$m bands only (Figure~{\ref{irac12}}). It is classified as a `Ture' PN in the MASH~II catalogue. The NVSS survey gives a flux density of $2.1\pm0.5$\,mJy at 1.4\,GHz \citep{condon98}.  Its H$\alpha$ image shows a seemingly bipolar structure, which however is hard to tell from its fuzzy 8\,$\mu$m image.

PNG 002.2$+00.5$.---    This source was detected by IRAC in all the four
bands (Figure~{\ref{irac13}}).
The NVSS survey gives a flux density of $15.2\pm0.7$\,mJy at 1.4\,GHz 
\citep{condon98}. The H$\alpha$ image exhibits an elliptical appearance
with resolved internal structure. The IRAC data show that it is
a very red PN and has a morphology visibly similar to that of the
archetypal PN NGC\,7027, i.e., a limb-brightened, bi-conical shell
with enhanced emission at the equator 
\citep[see][for the details of the image of NGC\,7027]{cox02}.

PNG 002.2$-01.2$.---    This source was detected by IRAC in the 4.5, 5.8, and 8.0\,$\mu$m bands only (Figure~{\ref{irac14}}). It is classified as a `Ture' PN by \citet{parker06}. Strong [\ion{N}{2}] and weak [\ion{O}{3}] lines were detected by \citet{parker06}, suggesting that it is a low excitation PN.
The NVSS survey gives a flux density of $2.1\pm0.5$\,mJy at 1.4\,GHz \citep{condon98}.  The 8.0\,$\mu$m 
and H$\alpha$ appearances are similar to each other and show a compact and rectangular shaped structure.

PNG 003.4$+01.4$.---     This source is almost invisible in the
IRAC bands (Figure~{\ref{irac15}}). We only find an extremely faint
counterpart in 
the 8.0\,$\mu$m band. Its 24\,$\mu$m counterpart appears to be
a point source and is fainter compared to other GLIMPSE~II PNs. The object 
is described by \citet{parker06} as having faint, semi-circular arc around faint central star, and is classified as a `likely' PN. Albeit with faintness in optical and MIR bands, it has a sufficiently strong radio flux.  The NVSS survey gives a flux density of $15.2\pm0.7$\,mJy at 1.4\,GHz \citep{condon98}. Its faint 8.0\,$\mu$m image does not allow us to say much about its infrared structure.

PNG 003.5$-01.2$.--- This source was detected by IRAC in  all the four
bands (Figure~{\ref{irac16}}). 
\citet{parker06} detected strong [\ion{N}{2}] and weak [\ion{O}{3}] lines in the object and classified it as a `Possible' PN, possibly a symbiotic star containing a Mira variable.
The object has an {\it IRAS} counterpart, {\it IRAS} $17554-2628$.
The NVSS survey gives a flux density of $13.7\pm0.6$\,mJy at 1.4\,GHz
\citep{condon98}.  The IRAC and 24\,$\mu$m images exhibit a
very red and bright point source. The SED implies the presence
of dust with a temperature of $\sim300$\,K (Figure~\ref{sed}). Moreover, the
increase of flux at $\sim100$\,$\mu$m points to the existence of another dust component with even lower temperature.

PNG 003.5$+01.3$.---     This source was detected by IRAC in  all the four bands (Figure~{\ref{irac17}}). It is classified as a `True' PN in the MASH~II catalogue. The NVSS survey gives a flux density of $24.9\pm0.9$\,mJy at 1.4\,GHz \citep{condon98}. Its IRAC images appear to be a point source.
The SED shows a marked increase of flux from 6--12\,$\mu$m,  suggesting the presence of cold dust of $T<300$\,K (Figure~\ref{sed}).

PNG 003.6$-01.3$.---    This source was detected by IRAC in  all the four bands (Figure~{\ref{irac18}}). It is described by \citet{parker06} as a small, bright, round, and `True' PN with strong [\ion{O}{3}] and H$\alpha$ emission. The IRAC images exhibit a similar appearance with H$\alpha$ image although the former seem to be more compact. The SED shows an emission peak around $1.5$\,$\mu$m (Figure~\ref{sed}), probably pointing to the substantial contribution from photospheric and nebular bound-free emission.

PNG 003.7$+00.5$.---  This source is badly blended with a very bright star 
(Figure~{\ref{irac19}}) which saturates IRAC. Nevertheless, 
we can still find evidence showing extended emission in the northwest of this
bright field star.  The object is described by \citet{parker06} as a very small, 
isolated slightly oval 
nebula with asymmetric enhance opposing edges, and is classified as a `Likely' 
PN. The 24\,$\mu$m emission is seriously polluted by the bright field star,
and thus we only give the upper limit of the flux.

PNG 004.3$-01.4$.---   This source was detected by IRAC in  all the four
bands (Figure~{\ref{irac20}}). It is classified as a `True' PN by
\citet{parker06}.
The NVSS survey gives a flux density of $12.7\pm0.7$\,mJy at 1.4\,GHz 
\citep{condon98}. This object has a {\it IRAS} counterpart, {\it IRAS} 
$17582-2553$.  The MIPS 24 $\mu$m flux of 1.7\,Jy is in excellent agreement with the 
25\,$\mu$m flux (1.9\,Jy) given in the {\it IRAS} PSC. 
\citet{kis95} imaged this PN through narrowband filter centered at [\ion{S}{3}]
$\lambda$9532, which reveals a size of $\sim2''$. The IRAC and H$\alpha$ 
images display a compact and roughly round structure. The SED shows a steep increase of flux at 10\,$\mu$m (Figure~\ref{sed}), and suggests that the PN might have a dust component with a color temperature of $\sim300$\,K.

PNG 004.8$-01.1$.--- This source was detected by IRAC in all the four bands 
(Figure~{\ref{irac21}}). It is classified as a `True' PN by
\citet{parker06}.  This object has a {\it IRAS} counterpart, {\it IRAS}
$17581-2522$. The [\ion{S}{3}] $\lambda$9532 narrowband image suggests a size of $4''$ \citep{kis95}. The smaller size at [\ion{S}{3}] compared to those at MIR and H$\alpha$ bands is due to the fact
that the former only traces the inner high-ionization regions.
The NVSS survey gives a flux density of $16.2\pm3.1$\,mJy 
at 1.4\,GHz \citep{condon98}. The 8\,$\mu$m and H$\alpha$ images show
an oval appearance. The SED suggests the presence of cold dust 
with $T<150$\,K (Figure~\ref{sed}).

PNG 006.1$+00.8$.---  This source was detected by IRAC in  all the four
bands (Figure~{\ref{irac22}}). It is classified as a `True' PN by
\citet{parker06}. The 8\,$\mu$m and H$\alpha$ images are similar and
show a compact and round appearance. 

PNG 352.1$-00.0$.---   This source was detected by IRAC in  all the four
bands (Figure~{\ref{irac23}}). The H$\alpha$ image shows an arcuate
nebula around a star. Much more extended structures are revealed
by the IRAC images which show  a sharp edge to the east and a diffuse
tail trailing toward the west. Based on its morphology, it is likely to be an \ion{H}{2} region and not a PN. The cometary structure shown by the MIR images is commonly seen in compact and ultracompact \ion{H}{2} 
regions and implies the interaction between the nebula and the ISM.
We also find that the object is surrounded by very extended
($\sim10$\,arcmin across) fainter filaments. The 24\,$\mu$m emission is 
dominated by the central bright region and appears as a seemingly oval 
structure. This object is an example showing that these MIR data can 
help for identifying/rejecting PNs.
We note that there are two bright IR sources, {\it IRAS} $17221-3533$ and 2MASS $J17253388-3536005$ \citep[the latter might be a YSO;][]{felli02}, 
lying within the nebulosity.

PNG 352.8$-00.5$.---   This source was detected by IRAC in  all the four
bands (Figure~{\ref{irac24}}). It has a {\it IRAS} counterpart,
 {\it IRAS}  $17262-3511$ and is classified as a `True' PN
in the MASH~II catalogue.  The NVSS survey gives a flux density of 
$72.8\pm2.3$\,mJy at 1.4\,GHz \citep{condon98}.
We note that the PN lies within an overlap region 
between the GLIMPSE~I and GLIMPSE~II surveys and its IRAC images have
already investigated by \citet{ramos08}. We refer the reader to their
paper for further details of this PN. Its 24\,$\mu$m emission is very strong
and saturates the detector. The SED shows a smooth increase of flux from
1--50\,$\mu$m, and suggests the presence of cold
dust with $T=60$--100\,K (Figure~\ref{sed}).

PNG 353.9$+00.0$.---   This source was detected by IRAC in  all the four
bands (Figure~{\ref{irac25}}). Optical spectroscopy suggests a low excitation class and the object is classified as a `Possible' PN by \citet{parker06}.
The H$\alpha$ and MIR images appear as a point source. Unlike the
other PNs, this source has a IRAC color only slightly redder than stars, and
has a low SED slope ($\alpha_{\rm IRAC}=-0.91$).
The SED indicates to a color temperature of about 1000\,K (Figure~\ref{sed}),
or even higher if the interstellar extinction is non-negligible.
This value is significantly
higher than in  normal PNs. Consequently, this object is more likely
to be an emission-line star rather than a PN.

PNG 355.6$-01.4$.---   This source is almost overwhelmed by the bright
background emission and was detected by IRAC in the 8.0\,$\mu$m band only 
(Figure~{\ref{irac26}}). It is classified as a `True' PN by \citet{parker06}. 
The NVSS survey gives a flux density of $5.5\pm0.6$\,mJy at 1.4\,GHz 
\citep{condon98}. The 8\,$\mu$m image is visually consistent with
the H$\alpha$ appearance that shows a small bipolar-like structure.

PNG 355.6$+01.4$.---  This source was detected by IRAC in  all the four
bands (Figure~{\ref{irac27}}). It is classified as a `Likely' PN by 
\citet{parker06}. The spectroscopy suggests that it probably is a
very low excitation PN.  The NVSS survey gives a flux density of 
$8.2\pm0.6$\,mJy at 1.4\,GHz \citep{condon98}. The H$\alpha$ and
MIR images appear as a point source.
The SED implies the presence of cold dust with $T<100$\,K
(Figure~\ref{sed}).

PNG 356.0$-01.4$.---  This source was detected by IRAC in the
5.6 and 8.0\,$\mu$m bands only (Figure~{\ref{irac28}}). 
 It is classified as a `True' PN by \citet{parker06}.
  The NVSS survey gives a flux density of
$4.8\pm0.5$\,mJy at 1.4\,GHz \citep{condon98}. The H$\alpha$ and
IRAC images show that it is a faint, compact, and elliptical PN.

PNG 356.9$+00.9$.---  This source was detected by IRAC in  all the four
bands (Figure~{\ref{irac29}}). It is classified as a `True' PN by 
\citet{parker06}.   The NVSS survey gives a flux density of
$12.5\pm0.6$\,mJy at 1.4\,GHz \citep{condon98}. The H$\alpha$ 
image appears as a point source. However, extended emission is
clearly resolved in the IRAC images which exhibit a round structure. 
The PN is very red and bright 
in the MIR bands. The SED implies the presence of cold dust with 
$T<150$\,K (Figure~\ref{sed}).

PNG 357.4$-01.3$.---  This source was detected by IRAC in  all the four bands (Figure~{\ref{irac30}}).  Given its large size ($\sim5$\,arcmin), the object is unique in our sample.  \citet{parker01} suggested that it might be a highly evolved PN. As the extended emission region is very diffuse, we cannot obtain reliable integrated flux and can only roughly estimate the size. 
The optical image shows that it is a ring nebula around the Wolf-Rayet
(WR) star WR~101. \citet{cappa02} reported  the radio
image of this nebula at 1465\,MHz taken through the VLA and obtained
an ionized mass of about 230\,M$_\sun$. Therefore, this object 
is unlikely to be a PN, and is 
probably a WR nebula consisting of mass lost as well as interstellar material 
swept-up by the stellar wind from an evolved massive star. We
find that the H$\alpha$ and 24\,$\mu$m images have a high resemblance.
However, compared to that in H$\alpha$ and 24\,$\mu$m,
the ring structure is less well defined and fuzzier in the IRAC bands,
which also reveal more diffuse emission extending to the northwest.
The position of the bright clump in the southeast as shown 
by IRAC slightly differs from that of the H$\alpha$ counterpart
which appears to be closer to the ionized source. This suggests
that the nebulosity is optically thick to the ionizing photons, and 
probably the dust grains close to the ionized source
have been destructed by photodissociation of the central star.

PNG 357.5$+01.3$.---  This source was detected by IRAC in  all the four
bands (Figure~{\ref{irac31}}). It is classified as a `Likely' PN
by \citet{parker06}. The H$\alpha$ image appears as a point source.
Extended emission is clearly resolved in the IRAC images, which
show that it is a red and round nebula.

PNG 357.7$+01.4$.--- This source was detected by IRAC in  all the four
bands (Figure~{\ref{irac32}}). It is classified as a `True' PN
by \citet{parker06}. The H$\alpha$ image shows a slightly elongated structure.
The IRAC images only reveal a bright core and the object is the most compact PN in our sample.  

PNG 358.2$-01.1$.---  This source was detected by IRAC in  all the four
bands (Figure~{\ref{irac33}}). This PN has been discovered before
\citep[e.g.][]{cahn71,acker91}.   The NVSS survey gives a flux density of
$<2.5$\,mJy at 1.4\,GHz \citep{condon98}.
Using the energy-balance method,
\citet{pre89} obtained the temperature of the central star
to be 144,000\,K. \citet{maciel84} determined a distance toward this PN of 
1.1\,kpc. It is described by \citet{parker01} as a bright, compact,
possibly bipolar PN with arc features.
The IRAC images have a visually similar appearance with the
H$\alpha$ image and show arc features and fuzzy extended structure.

PNG 358.8$-00.0$.---   This source was detected by IRAC in  all the four
bands (Figure~{\ref{irac34}}). The NVSS survey gives a flux density of
$<2.5$\,mJy at 1.4\,GHz \citep{condon98}.  Although the object has a 
{\it IRAS} counterpart, {\it IRAS} $17395-2950$, we cannot find the flux
measurements in the {\it IRAS} PSC. \citet{volk91} presented the low-resolution 
{\it IRAS} spectrum and classified the PN into Group~I which represents
noisy or incomplete spectra. The H$\alpha$ image shows a small and fuzzy
nebula. It has a large $\alpha_{\rm IRAC}$ value, and is
redder than all the other GLIMPSE~II PNs.
The MIR images are remarkably extended and reveal intriguing structure.
The bright region has a C-shape.
We find that it has a large dust envelope with a cavity centered on the central star and diffuse emission extending to the west. Thus it is unlikely to be a PN.  The SED implies the presence of cold dust with
$T<150$\,K (Figure~\ref{sed}).  From the IRAC images, we can also see a bipolar nebulosity in the northwest to PNG 358.8$-00.0$. This object has no visible optical counterpart and is probably associated with a compact radio source discovered by \citet{gray93}.  From the morphology, we conjecture that it is likely to be a PN or compact \ion{H}{2} region highly obscured by dust extinction. Follow-up observations of this GLIMPSE-survey-discovered object are required to unveil its nature.

PNG 359.1$-01.7$.---  This source was detected by IRAC in  all the four bands (Figure~{\ref{irac35}}). This is a previously known PN \citep[e.g.][]{henize67,webster75}. A detailed optical spectrophotometric study has been recently presented by \citet{wang07}. The deduced chemical abundances suggest that it a Type~I PN.  The NVSS survey gives a flux density of $91.9\pm2.8$\,mJy at 1.4\,GHz \citep{condon98}. 
\citet{zhang95} derived a distance to this PN of 3.30\,kpc. Using a dynamical method,  \citet{gesicki07} obtained the mass of the central star of $\sim0.627$\,M$_\sun$.
The IRAC and H$\alpha$ images show a compact, bright, and round appearance.  However, \citet{gesicki03} found that this object has complicate  line profiles of \ion{H}{1}, [\ion{N}{2}], and [\ion{O}{3}],
 and suggested the presence a bipolar outflow. In the IRAC images, we do note that the brightest part is not located in the center of the nebula.  The SED indicates to a color temperature $T<300$\,K (Figure~\ref{sed}).

PNG 359.2$+01.2$.---   This source was detected by IRAC in  all the four bands (Figure~{\ref{irac36}}). It has been known as a typical bipolar PN \citep[e.g.][]{kohoutek82,ratag90,cs95} and has been suggested as 
 a symbiotic system \citep{corradi95}. The NVSS survey gives a flux density of $13.5\pm0.6$\,mJy at 1.4\,GHz \citep{condon98}.  It has a {\it IRAS}  counterpart, {\it IRAS} $17358-2854$.
The H$\alpha$ image shows a bright core with a highly collimated bipolar outflow. The compact core has also been detected by \citet{lee07} at radio wavelengths.  The IRAC images show a less extreme bipolar shape with a lower major/minor axis ratio compared to the H$\alpha$ image. We speculate that the PN has
a central dust torus therefore enhancing the IRAC emission near the center. Interestingly, the appearance revealed by the 24\,$\mu$m image is a bipolar nebulosity surrounded by an extended halo with a size of $30\arcsec\times42\arcsec$. The SED indicates to the presence of cold dust with $T<100$\,K (Figure~\ref{sed}).

PNG 359.3$-00.9$.---  This source was detected by IRAC in  all the four bands (Figure~{\ref{irac37}}). 
This is a well-known bipolar nebula Hb 5 and has been extensively studied \citep[see, e.g.][for recent literature summaries of this object]{pottasch07,montez09}.  Optical images show that this PN is composed of two closed-end bipolar lobes, a bright compact core, concentric rings and some filaments.  It is a high-excitation Type~I PN and suffers from heavy extinction.  The distance to this PN is not accurately known, with estimates ranging from 1--7\,kpc \citep{montez09}.  It is one of the four PNs that have hitherto been detected with X-ray emission, indicating the presence of shock-heated gas in the bipolar nebula \citep{montez09}. The NVSS survey gives a flux density of $179.5\pm5.4$\,mJy at 1.4\,GHz \citep{condon98}.  The IRAC images exhibit the bright compact core, which is so luminous that the 8\,$\mu$m band is saturated. Because of the brightness of the central core, the bipolar lobes are not obvious from the IRAC images. It is clear that in the IRAC wavelengths, the core-to-lobe flux-density ratio is larger than that at H$\alpha$.
We suggest that the 8\,$\mu$m emission is dominated by the dust thermal emission from a central torus.
The 24\,$\mu$m band, however, shows more extended emission even when deconvolved using a point source function. The 24-$\mu$m image suggests an extended cold dust envelope, corresponding to the $\sim100$\,K dust component seen in the SED (Figure~\ref{sed}).  The ISO spectrum of Hb 5 is plotted on the SED,  showing  that ISO flux measurements are in good agreement with other measurements.
We note that there are two strong high-ionization lines [\ion{Ne}{5}] 24\,$\mu$m and [\ion{O}{4}] 26\,$\mu$m
lying in the 24\,$\mu$m band, which contribute about 10 percent of the total
24\,$\mu$m flux and are partly responsible for the saturation of the central 
region in the 24\,$\mu$m  image.

\subsection{Nondetections}

The following MASH PNs have no obvious IRAC counterparts:
PNG $000.0-01.8$,
PNG $000.1+01.9$,
PNG $000.6-01.4$,
PNG $001.0-01.4$,
PNG $001.1-01.2$   
PNG $001.2+01.3$, 
PNG $001.5-01.6$, 
PNG $001.6+00.1$, 
PNG $001.6-00.6$, 
PNG $001.6+01.6$, 
PNG $001.9+01.9$, 
PNG $002.1+01.2$, 
PNG $002.4+01.1$, 
PNG $002.5+01.3$, 
PNG $002.7-01.4$,
PNG $004.0-00.4$, 
PNG $004.8-00.5$, 
PNG $005.5-00.8$,
PNG $007.2+00.0$,
PNG $009.9-00.2$, 
PNG $354.0-00.8$,
PNG $354.3+00.5$,
PNG $354.8-00.5$,
PNG $355.5-01.1$, 
PNG $355.9+00.7$,
PNG $357.3+01.3$, 
PNG $357.6+01.0$, 
PNG $359.1-00.7$, 
PNG $359.2-01.2$, 
PNG $359.2+01.3$, 
PNG $359.3+01.4$, 
PNG $359.5-01.8$, 
PNG $359.9-01.8$,
PNG $359.9+01.8$.
In most of the cases, the detection of these PNs is hampered by  the bright background near the Galactic center.  For several objects (e.g. PNG $001.1-01.2$ and PNG $002.7-01.4$), the IRAC bands exhibit apparent point sources near the corresponding positions, but without extended emission around these sources. They are likely to be field stars although some of them have a rising flux distribution from 3.8 to 8.0\,$\mu$m.  
An interesting object is PNG $354.8-00.5$, which is a radio and maser source.  Its H$\alpha$ image reveals a 
faint, circular nebula with a size of $42.0\arcsec\times37.0\arcsec$.  However, the IRAC images show an extremely luminous star, which is bright enough to saturate the detector and makes the nebula `invisible'. The object probably represent a evolutionary stage of PN formation around a very evolved AGB star, as suggested by \citet{cohen05b}.

\section{DISCUSSION}

All the objects studied in our sample can be categorized as ionized nebulae with cold dust emission.  The ionized nature of the objects is confirmed by H$\alpha$ and radio continuum emission.  The peak of the dust continuum mostly
lies between 10 and 30 $\mu$m, suggesting dust temperatures of 100--300 K.  
Since the infrared spectra of PNs are rich in ionic emission lines as well as solid-state emission bands \citep[see, e.g.,][]{sta07}, 
accurate determination of dust temperatures of PNs is very difficult without 
spectra and our estimates represent only rough approximations.

\subsection{Implications for Source Sizes at Different Wavelengths}

It is instructive to compare the morphologies observed in different
bands since they could trace the spatial distributions of
different nebular components.
Figure~\ref{size} shows histograms of source sizes in MIR 
wavelengths relative to H$\alpha$, for which we have deconvolved the point 
spread functions (PSF) simply through 
$\theta_{\rm s}=\sqrt{\theta_{\rm obs}^2-\theta_{\rm res}^2}$
(note that this is only a rough estimate  as the PSFs are non-Gaussian),
where $\theta_{\rm s}$ is the intrinsic source sizes,
$\theta_{\rm obs}$ is the observed sizes, and $\theta_{\rm res}$ is the angular resolution. An
inspection of this figure shows that (1) the H$\alpha$ sizes are more similar to those observed in 5.8 and 8.0\,$\mu$m compared to the other MIR sizes, (2) the 24\,$\mu$m sizes are generally
larger than those in the other wavelengths, and (3) The 3.6 and 4.5\,$\mu$m sizes are quite similar to each other and are more compact than those in the other bands.

The optical and 8.0\,$\mu$m morphologies of a certain PN can be remarkably different. For some PNs, the 8.0\,$\mu$m emission has a larger angular extent than the corresponding optical image. This is indicative of the presence of extended dust envelopes which obscure the optical emission and emit thermal continuum and AIB features contributing to the 8.0\,$\mu$m band. This can also in part be attributed to the H$_2$ rotational-vibrational transitions excited by shocks.  On the other hand, some PNs are more compact at 8\,$\mu$m than at H$\alpha$. An important issue is whether this really reflects the different spatial distributions of the different components rather than different instrumental sensitivities. In order to investigate this problem, in Figure.~\ref{profile} we compare the H$\alpha$ and 8.0\,$\mu$m flux profiles along the major axis of the bipolar PN PNG $359.2+01.2$. While the H$\alpha$ and 8.0\,$\mu$m images have comparable angular resolutions, it is clear that the 8.0\,$\mu$m emission is more centrally concentrated than the H$\alpha$ emission.  The most likely interpretation is that the H$\alpha$ and 8.0\,$\mu$m emissions originate from different regions, the former from the bipolar lobes and the latter from a nuclear dust torus. Examples of infrared-observed equatorial dust torus can be found in the bipolar proto-PN IRAS $17441-2411$ \citep{volk07}. If we generalize from this example, objects with smaller angular extents in 8.0\,$\mu$m than in H$\alpha$ may be bipolar nebulae.

The  24\,$\mu$m images of our sample objects are in general well resolved, extended, and show a circular appearance with a sharp boundary.  It is unlikely that [\ion{Ne}{5}] 24.3\,$\mu$m and [\ion{O}{4}] 25.9\,$\mu$m lines are dominant contributor to the 24\,$\mu$m flux otherwise the 24\,$\mu$m image would show a 
more compact appearance than
the H$\alpha$ image.  It is more probable that the 24\,$\mu$m images correspond to the distribution of cold dust which is the remnant of AGB mass loss.
Extended AGB halos have been frequently detected in deep optical images of
PNs \citep{corradi03}.
If the cold dust grains are plentifully present within these halos,
they can exhibit strong 24 $\mu$m emission.


For high-excitation PNs, the [\ion{Ne}{5}] and [\ion{O}{4}] line emission may significantly enhance the 24\,$\mu$m emission in nebular centers.  A recent study of MIPS  24\,$\mu$m images of Galactic PNs (which are generally more extended than objects in our sample) has been reported by \citet{chu09}, who find that the 
sizes and surface brightnesses of H$\alpha$ and 24\,$\mu$m emission suggest an evolutionary sequence.

From Figures~\ref{mips}--\ref{mipsend}, we can see some of our sample objects show faint and extended halos and ring structures around their bright cores.  While some of these are clearly instrumental artifacts,  some are likely to be real (e.g., PNG $001.0-01.9$, PNG $001.0+01.9$, PNG $359.1-01.7$ etc.) and deserve further study.  A recent MIPS24-based study to search for debris disk around stars has found 24\,$\mu$m excess emission in the Helix  Nebula and suggested that it originates from a dust disk at a distance
of 40--100\,AU from the central star, and that the dust is produced by collisions of Kuiper-Belt-like objects
 \citep{su07}.  Assuming that the distances to the GLIMPSE~II PNs are typically larger than 1\,kpc, we find that the 24\,$\mu$m  emission extends to $>5000$\,AU from the center, more than a factor of 100 times farther than the Sun's Kuiper Belt, and thus is unlikely to arise from a Kuiper Belt around low- and
intermediate-mass AGB stars. However, we cannot completely rule out the possibility that Kuiper Belt-like objects may partly contribute to the 24\,$\mu$m emission in the centers.  An interesting PN is PNG $001.0+01.9$ which is the only object showing more compact 24\,$\mu$m  emission relative to H$\alpha$ in our sample. Its 24\,$\mu$m emission is probably in part associated with the central star.

For the two short-wavelengths IRAC bands (3.6 and 4.5\,$\mu$m), the contributions from the hydrogen recombination line and bound-free emission should not be dominant, because if so, the 3.6 and 4.5\,$\mu$m emission would 
morphologically match the H$\alpha$ emission well.  The spatially compact 3.6 and 4.5\,$\mu$m emission  probably suggest significant contribution from photospheric emission. This conclusion is further supported by the SEDs (Figure~\ref{sed}) and color-color diagrams (Figures~\ref{color}
and \ref{color2}).

\subsection{IRAC 8.0\,$\mu$m versus {\it MSX} 8.3\,$\mu$m Integrated Fluxes}

\citet{reach05} investigated the absolute calibration of IRAC.
They found that the extended emission surface brightness is incorrectly
calibrated and should be multiplied by an effective aperture correction
factor (0.737 for the 8.0\,$\mu$m band). 
This was subsequently established by \citet{cohen07a} that the IRAC 8.0\,$\mu$m surface brightness should be scaled down by a factor of $0.74\pm0.07$ based on comparison between the IRAC 8.0\,$\mu$m and {\it MSX} 8.3\,$\mu$m fluxes of a sample of 43 \ion{H}{2} regions.  This study predicts a IRAC8.0/{\it MSX}8.3 flux ratio of $1.14\pm0.02$ for diffuse \ion{H}{2} regions based on the different contributions from AIB emission in the two different bandpasses.  
A similar study of 14 PNs \citep{cohen07b} also confirm their previous results.  Here we present an independent examination of the IRAC 8.0\,$\mu$m calibration using the integrated fluxes of
the GLIMPSE~II PNs which are very compact diffuse sources.

A total of 10 GLIMPSE~II PNs have {\it MSX} 8.3\,$\mu$m detections.
Figure~\ref{compar1} compared the spatially integrated fluxes
detected in the IRAC 8.0\,$\mu$m and {\it MSX} 8.3\,$\mu$m bands.
For comparison, we also overplot the values for GLIMPSE~I PNs.
A good correlation between the two fluxes is clearly shown.
The PN with the largest IRAC/{\it MSX} flux discrepancy is PNG $003.5+01.3$, 
which has a large {\it MSX} 8.3\,$\mu$m flux compared with
the  IRAC 8.0\,$\mu$m flux.
Combining the GLIMPSE~I and GLIMPSE II measurements,
we deduce an average IRAC8.0/{\it MSX}8.3 flux ratio
of $0.90\pm0.36$, consistent with the median value
of $1.2\pm0.2$ found by \citet{cohen07b} for PNs but slightly
lower than that of $1.55\pm0.15$ obtained by \citet{cohen07a} for 
\ion{H}{2} regions.

\citet{cohen07b} found a marginal relation between the IRAC8.0/{\it MSX}8.3 
flux ratio and nebular diameters which suggests that the aperture correction
factor increases from 1.0 at small nebular size (i.e. point-source) to 
$\ge1.4$ at $77\arcsec$. This relation is not shown in Figure~\ref{compar1} 
because our sample does not include very extended PNs.
The PNs investigated here have a diameter smaller than 60$\arcsec$,
and mostly smaller than  20$\arcsec$, much more compact than
the arcminutes-scale \ion{H}{2} regions. Therefore, we infer that the absolute
flux calibration for the compact sources studied here should
be accurate, and we do not need to make aperture correction to obtain
the integrated fluxes of the GLIMPSE~II PNs.

The [\ion{S}{4}] 10.5\,$\mu$m line, which can contribute to the 
{\it MSX} 8.3\,$\mu$m band but not to the IRAC 8.0\,$\mu$m band,
 is partly responsible to the IRAC8.0/{\it MSX}8.3 flux ratio. PNs often have
higher excitation class than \ion{H}{2} regions, and thus are supposed
to have stronger [\ion{S}{4}]  emission. As a result, the
IRAC8.0/{\it MSX}8.3 flux ratio in PNs (especially high-excitation PNs) 
might be lower compared with that in \ion{H}{2} regions.  

We cannot completely rule out variability as the partial cause
of the IRAC and {\it MSX}  flux difference  for individual PNs
since the GLIMPSE survey was performed
about ten years after the {\it MSX} observations. If this is
the case, our study will have implications to understand the pulsations 
mechanism 
or binary system in PNs.  Another possible reason causing the IRAC/{\it MSX} flux
discrepancy in individual PNs is that the {\it MSX} fluxes of some PNs might be 
contaminated by blended sources since {\it MSX} has a resolution approximate ten times lower than IRAC.

\subsection{MIPS 24\,$\mu$m versus {\it MSX} 21\,$\mu$m Integrated Fluxes}

The ratio of  MIPS 24\,$\mu$m to {\it MSX} 21\,$\mu$m  fluxes
provides an opportunity to establish the reliability of absolute diffuse 
calibration and our measurements of MIPS 24\,$\mu$m fluxes. For this
purpose, one should at first estimate the predicted MIPS24/{\it MSX}21 flux 
ratio  through integrating the spectral response functions to
the SEDs. \citet{cohen09} investigated the diffuse absolute 
calibration of the MIPS 24\,$\mu$m channel using a sample of \ion{H}{2} regions,
and obtained the observed MIPS24/{\it MSX}21 flux ratio of
$1.19\pm0.03$. Comparing with the predicted value of $1.18\pm0.02$,
which is  obtained from three SEDs pertinent to  \ion{H}{2} regions,
they concluded that the absolute diffuse calibration of MIPS24 is consistent 
with that of {\it MSX}21 within $3\%$ ($1\sigma$). 
We reasonably suppose that predicted MIPS24/{\it MSX}21 flux ratio of PNs
is nearly the same as that of \ion{H}{2} regions.

In Figure~\ref{compar2} we compare the MIPS 24\,$\mu$m and
{\it MSX} 21\,$\mu$m integrated fluxes of the GLIMPSE~II PNs. The 
sensitivity of the {\it MSX}21 band is approximately 1000 times lower than
of the MIPS24 band, and thus only relatively bright PNs can be detected
by {\it MSX}21.
We found that nine of the GLIMPSE~II PNs have {\it MSX}21 counterparts.
Since the MIPS24 band has a much lower
saturated level, very bright sources in the {\it MSX} catalogue could be
saturated in MIPS. This is clearly reflected in Figure~\ref{compar2}.
Excluding the two saturated sources, we obtain the MIPS24/{\it MSX}21 
flux ratio of $1.19\pm0.12$, in perfect agreement with value for
\ion{H}{2} regions \citep{cohen09}. 
We note that the \ion{H}{2} regions
investigated by \citet{cohen09} are much more extended,
and generally have 24\,$\mu$m integrated
fluxes ranging from 10$^{4.0}$--10$^{5.5}$\,mJy, significantly
larger than those of the GLIMPSE~II PNs in Figure~\ref{compar2}
(about 10$^{3.2}$--10$^{3.9}$\,mJy). The consistent results
obtained from PNs and \ion{H}{2} regions, which have
remarkably different spatial sizes and integrated fluxes, strongly validate
the reliabilities of the absolute diffuse calibration of MIPS24
and our flux measurements.

\subsection{Comparison with Radio Flux Densities}

A tight correlation between far- and mid-infrared and radio emission from star-forming galaxies has been established and has been a subject of intense discussion for many years \citep[see, e.g.,][and the references therein]{appleton04}. The origin of the tightness of this correlation is, however, not 
completely understood. It would be interesting to test the MIR/radio relation 
in PNs, although they have quite different physical conditions from star
formation regions.

If the MIR emission is dominated by dust heated by ultraviolet (UV) photons from the central stars which also
ionize the gaseous nebulae and are responsible for free-free thermal emission in the  radio continuum,  we 
would see a correlation between the MIR and radio emission in PNs.
\citet{cohen07b} compared the IRAC 8.0\,$\mu$m, {\it MSX} 8.3\,$\mu$m, and 
radio flux densities and obtained an overall MIR/radio ratio of $4.6\pm1.2$.
They also suggest that, unlike that in \ion{H}{2} regions, the MIR/radio ratio in PNs might change with PN evolution as the result of decreasing amount of UV photons available to pump the AIB  into
MIR fluorescence.  Even without considering the line emission factor, the dust continuum flux of PNs will decrease as the dust cools due to geometric dilution of the heating photons, which will occur at a different rate as the radio continuum flux which decreases as the electron density drops as the result of expansion even the electron temperature remains constant.

We note that all the GLIMPSE~II PNs are located within the NVSS 1.4\,GHz survey \citep{condon98} field, and 23 of them have NVSS detections. They have sizes smaller than the 1.4\,GHz beam size of VLA and are spatially unresolved at this frequency. In Figure~\ref{radio} we compare the IRAC 8.0\,$\mu$m and
MIPS 24\,$\mu$m integrated fluxes with those from NVSS catalogue.  It is evident that albeit with a large dispersion, there is a positive correlation between the MIR and radio emission. The correlation
coefficients in the logarithmic space are 0.66 for IRAC 8.0\,$\mu$m vs 
NVSS 1.4\,GHz, and 0.64 for MIPS 24\,$\mu$m vs NVSS 1.4\,GHz.
Figure~\ref{radio} also shows that the number distributions of the
IRAC8.0$\mu$m/1.4GHz and MIPS24$\mu$m/1.4GHz flux ratios are peaked at
$\sim3$ and $\sim100$, respectively. If we assume that these GLIMPSE~II
PNs are approximately at the same evolutionary stage and that
the non-thermal emission is insignificant, the dispersion in Figure~\ref{radio} should be due to self-absorption of radio emission.  For most compact PNs, the turnover (from optically thick to optically thin) frequency is usually at 1-5 GHz, so it is quite possible that some of these nebulae are optically thick at 1.4 GHz.
The object showing the largest MIR/radio flux ratio is PNG $359.2+01.2$ which has a high density torus.

A basic observable parameter $q_{\rm IR}  [=\log(S_{\rm IR}/S_{1.4\rm GHz})$, $S$ being the flux density] is often used to characterize the MIR/radio relation.
For instance, \citet{appleton04} obtained  $q_{24}=1.00\pm0.27$ for galaxies with different redshifts;
\citet{ibar08} found $q_{24}=0.71\pm0.47$ for a sample at high-redshifts;
\citet{wu08} obtained the $q_{24}$ values for dwarf star-forming galaxies generally ranging from  0.9 to 1.3 with one extreme case of 2.2. For the GLIMPSE~II PNs, we obtain $q_{8.0}=0.79\pm0.52$ and $q_{24}=2.13\pm0.37$.  The $q_{8.0}$ value is consistent with that found for PNs in the Magellanic Clouds by \citet{filipovic09}.
The larger $q_{24}$ value in PNs compared to those in star-forming galaxies is a reflection of the fact that PNs and galaxies have different radio and infrared emitting mechanisms.  Galaxies radiate non-thermal radiation whereas PNs radiate thermal f-f emission.  The dust components in PNs are heated by a single central star whereas 
the dust in galaxies is heated primarily by massive stars.


\section{CONCLUSIONS}

An investigation of the MASH PNs within the GLIMPSE~II field has revealed that 37 optically identified PNs have MIR counterparts. These PNs are distinguishable from the surrounding stars by their redder color and more extended angular sizes. From their infrared morphologies, we suggest that some of these objects (PNG 352.1-00.0, PNG 353.9+00.0, PNG 357.4-01.3, PNG 358.8-00.0) are unlikely to be PNs.  We present the spatially integrated flux and size measurements of these PNs at 3.6, 4.5, 5.8, 8.0, and 24\,$\mu$m. MIR color-color and color-magnitude diagrams are produced.
Combining our results with those in other data archives, we construct the 0.82--100\,$\mu$m SEDs for these objects. The primary findings from analysis of these results are  summarized below.

\begin{list}{}{}

\item[1.]  Compared to the GLIMPSE~I PNs, the GLIMPSE~II PNs are generally more compact and have lower spatially integrated fluxes, implying that these PNs towards the Galactic bulge might be generally more distant. Nevertheless, the MIR colors of GLIMPSE~I and II PNs are very similar to each other, and are clearly different with those of stellar objects.

\item[2.] Despite some GLIMPSE~II PNs are invisible in the 3.6--5.8\,$\mu$m bands and are only marginally above the IRAC 8.0\,$\mu$m  detection limitation, all of them are characterized with very strong 24\,$\mu$m emission, which is likely to originate from extended cold dust envelopes.

\item[3.] The nebular morphologies and fluxes detected in different wavelengths allow us to separate the different components in the nebulae.  
The 5.8 and 8.0\,$\mu$m emission may be either more or less compact than their optical counterparts, respectively
implying the presence of dust toruses centered on bipolar PNs or extended neutral envelopes around ionized gaseous nebulae.

\item[4.] From the anomalous colors of some objects, we infer the contributions of AIB emissions to the IRAC fluxes.

\item[5.] The absolute diffuse calibrations are examined by comparing the IRAC 8.0\,$\mu$m and MIPS 24\,$\mu$m to the {\it MSX} 8.3 and 21\,$\mu$m fluxes. We obtain the ${\rm IRAC}8.0/MSX8.3$ and ${\rm MIPS}24/MSX21$ flux ratios of $0.90\pm0.36$ and $1.19\pm0.12$, respectively. These results are reasonably consistent with previous studies
by \citet{cohen07a}, \citet{cohen07b}, and \citet{cohen09} using data for \ion{H}{2} regions and GLIMPSE~I PNs, and suggest good calibrations for these slightly resolved GLIMPSE~II PNs.

\item[6.] We explore the MIR/radio relation for PNs. A loose correlation between the MIR (8.0\,$\mu$m and 24\,$\mu$m) and the NVSS 1.4\,GHz fluxes is found, and can be statistically described by $q_{\rm IR}$---the ratio between the flux densities at MIR and 1.4\,GHz. For the GLIMPSE~II PNs, we find  $q_{8.0}=0.79\pm0.52$ and $q_{24}=2.13\pm0.37$.  The larger $q_{\rm IR}$ values and the larger dispersion in the MIR/radio 
correlation of PNs compared to those of star-forming galaxies are attributed to their respective different underlying radiation mechanisms.

\end{list}

Since the bulk of GLIMPSE~II PNs are located at higher Galactic 
latitudes ($1^\circ\le|b|\le2^\circ$), we anticipate that a large 
number of PNs might have been detected in the GLIMPSE~3D legacy survey,
which extends the latitude coverage to $\pm3^\circ$.
The study of GLIMPSE~3D PNs will be presented in a subsequent paper.

\acknowledgments

We thank Ed Churchwell and the GLIMPSE team for assistance in the processing and analysis of GLIMPSE survey data.   We also thank Nico Koning for his help in data processing and Jun-ichi Nakashima for fruitful discussion.  This work is based on observations made with the {\it
Spitzer Space Telescope}, which is operated by the Jet Propulsion Laboratory, California Institute of 
Technology, under a contract with NASA.  This study used data products from the GLIMPSE and MIPSGAL 
{\it Spitzer Space Telescope} Legacy Programs.  This publication also used the data products from the DENIS project, which has been partly funded by the SCIENCE and the HCM plans of the European  Commission under grants CT920791 and CT940627, from the Two Micron All Sky Survey, which is a joint project of the University of Massachusetts and the Infrared Processing and Analysis Center/California Institute of Technology,
funded by the National Aeronautics and Space Administration and the National
Science Foundation, and from the Midcourse Space
Experiment, which was funded by the Ballistic Missile Defense Organization
with additional support from NASA Office of Space Science.
This research has also made use of the NASA/IPAC Infrared Science Archive, 
which is operated by the Jet Propulsion Laboratory, California Institute of 
Technology, under contract with the National Aeronautics and Space 
Administration, and made use of the SIMBAD data base, operated at CDS,
Strasbourg, France. Support for this work was provided by the Research 
Grants Council of the Hong Kong under grants HKU7032/09P.

{\rotate
\begin{figure*}
\epsfig{file=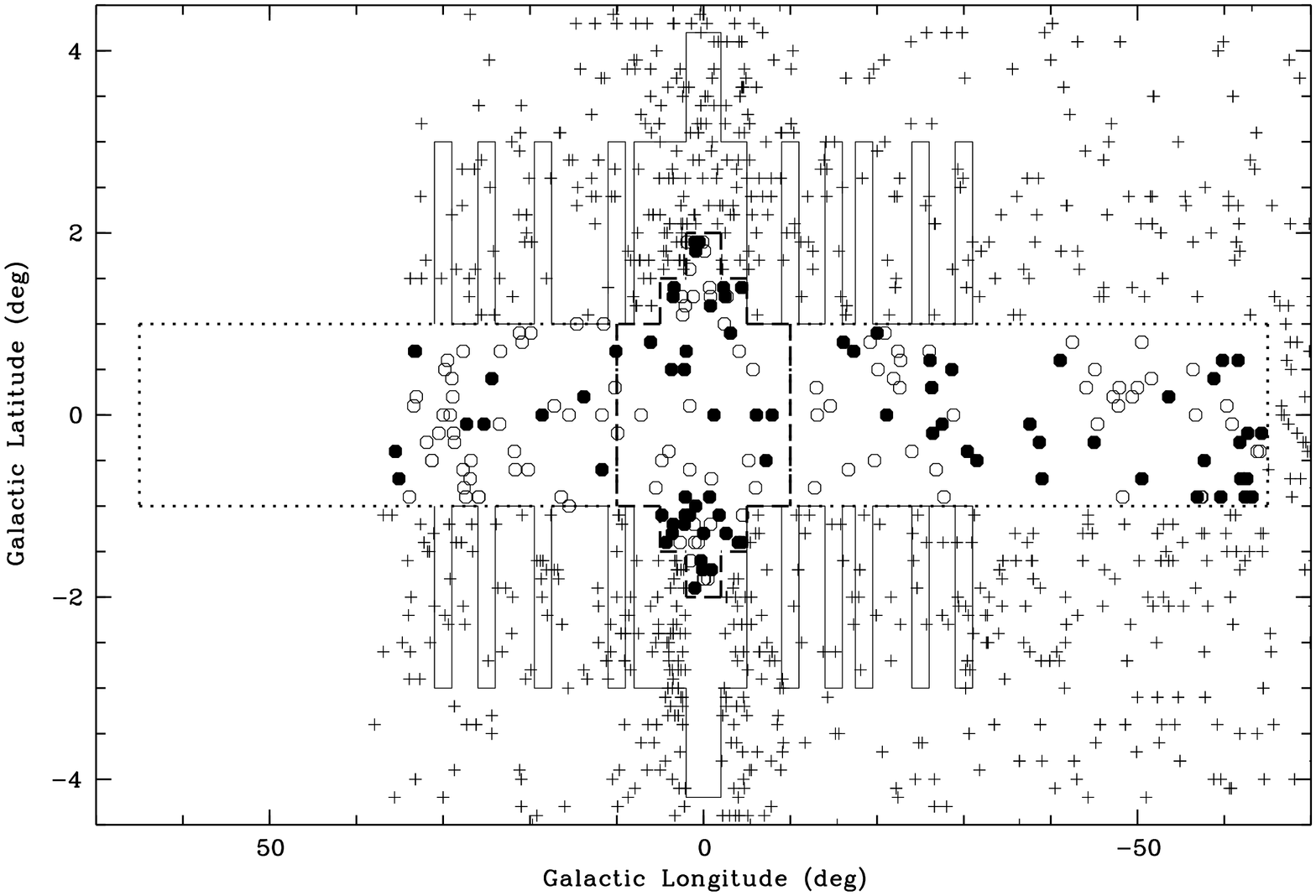, height=19cm,
bbllx=37, bblly=32, bburx=553, bbury=778, clip=, angle=-90}
\caption{  The survey coverage of GLIMPSE~I (dotted lines), GLIMPSE~II (dotted-dashed lines), and GLIMPSE~3D (solid lines). The circles and pluses represent the MASH PNs inside and outside the GLIMPSE~I/II survey coverage, respectively. The filled circles denote the MASH PNs having IRAC counterparts that are presented
in this paper and Paper~I. }
\protect\label{lb}
\end{figure*}
}


\clearpage

\newpage

\begin{figure*}
 \centering
 \caption{Composite-color images of PNG $000.0-01.3$ made from 3 IRAC bands.  The 3.6\,$\mu$m is shown as blue, the 5.8\,$\mu$m is shown as green, and the 8.0\,$\mu$m is shown  as red. The lower panel is the same with the upper panel but overlaid with the contours of H$\alpha$ emission. The target is  put in the center of each panel. The coordinates are in units of R.A. (J2000.0 on the horizontal scale) and decl.
(J2000.0 on the vertical scale). }
 \label{irac} 
\end{figure*}
\begin{figure*}
 \centering
 \caption{As in Figure.~\ref{irac}, but for PNG $000.1-01.7$.
 }
 \label{irac2} 
\end{figure*}
\begin{figure*}
 \centering
 \caption{As in Figure.~\ref{irac}, but for PNG $000.3-01.6$.
 }
 \label{irac3} 
\end{figure*}

\begin{figure*}
 \centering
 \caption{As in Figure.~\ref{irac}, but for PNG $000.5+01.9$.
 }
 \label{irac4} 
\end{figure*}

\begin{figure*}
 \centering
 \caption{As in Figure.~\ref{irac}, but for PNG $000.9-01.0$.
 }
 \label{irac5} 
\end{figure*}

\begin{figure*}
 \centering
 \caption{As in Figure.~\ref{irac}, but for PNG $000.9+01.8$.
 }
 \label{irac6} 
\end{figure*}

\begin{figure*}
 \centering
 \caption{As in Figure.~\ref{irac}, but for PNG $001.0-01.9$.
 }
 \label{irac7} 
\end{figure*}

\begin{figure*}
 \centering
 \caption{As in Figure.~\ref{irac}, but for PNG $001.0+01.9$.
 }
 \label{irac8} 
\end{figure*}

\begin{figure*}
 \centering
 \caption{As in Figure.~\ref{irac}, but for PNG $001.6-01.1$.
 }
 \label{irac9} 
\end{figure*}

\begin{figure*}
 \centering
 \caption{As in Figure.~\ref{irac}, but for PNG $002.0+00.7$.
 }
 \label{irac10} 
\end{figure*}

\begin{figure*}
 \centering
 \caption{As in Figure.~\ref{irac}, but for PNG $002.1-00.9$.
 }
 \label{irac11} 
\end{figure*}

\begin{figure*}
 \centering
 \caption{As in Figure.~\ref{irac}, but for PNG $002.1-01.1$.
 }
 \label{irac12} 
\end{figure*}

\begin{figure*}
 \centering
 \caption{As in Figure.~\ref{irac}, but for PNG $002.2+00.5$.
 }
 \label{irac13} 
\end{figure*}

\begin{figure*}
 \centering
 \caption{As in Figure.~\ref{irac}, but for PNG $002.2-01.2$.
 }
 \label{irac14} 
\end{figure*}

\begin{figure*}
 \centering
 \caption{As in Figure.~\ref{irac}, but for PNG $003.4+01.4$.
 }
 \label{irac15} 
\end{figure*}

\begin{figure*}
 \centering
 \caption{As in Figure.~\ref{irac}, but for PNG $003.5-01.2$.
 }
 \label{irac16} 
\end{figure*}

\begin{figure*}
 \centering
 \caption{As in Figure.~\ref{irac}, but for PNG $003.5+01.3$.
 }
 \label{irac17} 
\end{figure*}

\clearpage

\begin{figure*}
 \centering
 \caption{As in Figure.~\ref{irac}, but for PNG $003.6-01.3$.
 }
 \label{irac18} 
\end{figure*}

\begin{figure*}
 \centering
 \caption{As in Figure.~\ref{irac}, but for PNG $003.7+00.5$.
 }
 \label{irac19} 
\end{figure*}

\begin{figure*}
 \centering
 \caption{As in Figure.~\ref{irac}, but for PNG $004.3-01.4$.
 }
 \label{irac20} 
\end{figure*}

\begin{figure*}
 \centering
 \caption{As in Figure.~\ref{irac}, but for PNG $004.8-01.1$.
 }
 \label{irac21} 
\end{figure*}

\begin{figure*}
 \centering
 \caption{As in Figure.~\ref{irac}, but for PNG $006.1+00.8$.
 }
 \label{irac22} 
\end{figure*}

\begin{figure*}
 \centering
 \caption{As in Figure.~\ref{irac}, but for PNG $352.1-00.0$.
 }
 \label{irac23} 
\end{figure*}

\begin{figure*}
 \centering
 \caption{As in Figure.~\ref{irac}, but for PNG $352.8-00.5$.
 }
 \label{irac24} 
\end{figure*}

\begin{figure*}
 \centering
 \caption{As in Figure.~\ref{irac}, but for PNG $353.9+00.0$.
 }
 \label{irac25} 
\end{figure*}

\begin{figure*}
 \centering
 \caption{As in Figure.~\ref{irac}, but for PNG $355.6-01.4$.
 }
 \label{irac26} 
\end{figure*}

\begin{figure*}
 \centering
 \caption{As in Figure.~\ref{irac}, but for PNG $355.6+01.4$.
 }
 \label{irac27} 
\end{figure*}

\begin{figure*}
 \centering
 \caption{As in Figure.~\ref{irac}, but for PNG $356.0-01.4$.
 }
 \label{irac28} 
\end{figure*}

\begin{figure*}
 \centering
 \caption{As in Figure.~\ref{irac}, but for PNG $356.9+00.9$.
 }
 \label{irac29} 
\end{figure*}

\begin{figure*}
 \centering
 \caption{As in Figure.~\ref{irac}, but for PNG $357.4-01.3$.
 }
 \label{irac30} 
\end{figure*}

\begin{figure*}
 \centering
 \caption{As in Figure.~\ref{irac}, but for PNG $357.5+01.3$.
 }
 \label{irac31} 
\end{figure*}

\begin{figure*}
 \centering
 \caption{As in Figure.~\ref{irac}, but for PNG $357.7+01.4$.
 }
 \label{irac32} 
\end{figure*}

\begin{figure*}
 \centering
 \caption{As in Figure.~\ref{irac}, but for PNG $358.2-01.1$
 }
 \label{irac33} 
\end{figure*}

\begin{figure*}
 \centering
 \caption{As in Figure.~\ref{irac}, but for PNG $358.8-00.0$.
 }
 \label{irac34} 
\end{figure*}

\begin{figure*}
 \centering
 \caption{As in Figure.~\ref{irac}, but for PNG $359.1-01.7$.
 }
 \label{irac35} 
\end{figure*}

\clearpage

\begin{figure*}
 \centering
 \caption{As in Figure.~\ref{irac}, but for PNG $359.2+01.2$.
 }
 \label{irac36} 
\end{figure*}

\begin{figure*}
 \centering
 \caption{As in Figure.~\ref{irac}, but for PNG $359.3-00.9$.
 }
 \label{irac37} 
\end{figure*}


\begin{figure*}
\centering
\begin{tabular}{cc}
\end{tabular}
\caption{MIPS 24$\mu$m images of the GLIMPSE~II PNs.
From left to right:
PNG $000.0-01.3$ and PNG $000.1-01.7$ (1st row);
PNG $000.3-01.6$ and PNG $000.5+01.9$ (2nd row);
PNG $000.9-01.0$ and PNG $000.9+01.8$ (3rd row);
PNG $001.0-01.9$ and PNG $001.0+01.9$ (4th row).
The target is put in the center of each panel.
The coordinates are in units of R.A. (J2000.0 on the horizontal scale)
and decl. (J2000.0 on the vertical scale).}
\label{mips}
\end{figure*}

\begin{figure*}
\centering
\begin{tabular}{cc}
\end{tabular}
\caption{ MIPS 24$\mu$m images of the GLIMPSE~II PNs.
From left to right:
PNG $001.6-01.1$ and PNG $002.0+00.7$ (1st row);
PNG $002.1-00.9$ and PNG $002.1-01.1$ (2nd row);
PNG $002.2+00.5$ and PNG $002.2-01.2$ (3rd row);
PNG $003.4+01.4$ and PNG $003.5-01.2$ (4th row).
Details are indicated in Figure.~\ref{mips}. }
\end{figure*}

\newpage

\begin{figure*}
\centering
\begin{tabular}{cc}
\end{tabular}
\caption{ MIPS 24$\mu$m images of the GLIMPSE~II PNs.
From left to right:
PNG $003.5+01.3$ and PNG $003.6-01.3$ (1st row);
PNG $003.7+00.5$ and PNG $004.3-01.4$ (2nd row);
PNG $004.8-01.1$ and PNG $006.1+00.8$ (3rd row),
PNG $352.1-00.0$ and PNG $352.8-00.5$ (4th row).
Details are indicated in Figure.~\ref{mips}. }
\end{figure*}

\begin{figure*}
\centering
\begin{tabular}{cc}
\end{tabular}
\caption{ MIPS 24$\mu$m images of the GLIMPSE~II PNs.
From left to right:
PNG $353.9+00.0$ and PNG $355.6-01.4$ (1st row);
PNG $355.6+01.4$ and PNG $356.0-01.4$ (2nd row);
PNG $356.9+00.9$ and PNG $357.4-01.3$ (3rd row);
PNG $357.5+01.3$ and PNG $357.7+01.4$ (4th row).
Details are indicated in Figure.~\ref{mips}. }
\end{figure*}

\begin{figure*}
\centering
\begin{tabular}{cc}
\\
\end{tabular}
\caption{ MIPS 24$\mu$m images of the GLIMPSE~II PNs.
From left to right:
PNG 358.2-01.1 and PNG 358.8-00.0 (1st row); 
PNG 359.1-01.7 and PNG 359.2+01.2 (2nd row); 
PNG 359.3-00.9 (3rd row).
Details are indicated in Figure.~\ref{mips}. }
\label{mipsend}
\end{figure*}

\begin{figure*}
\centering
\epsfig{file=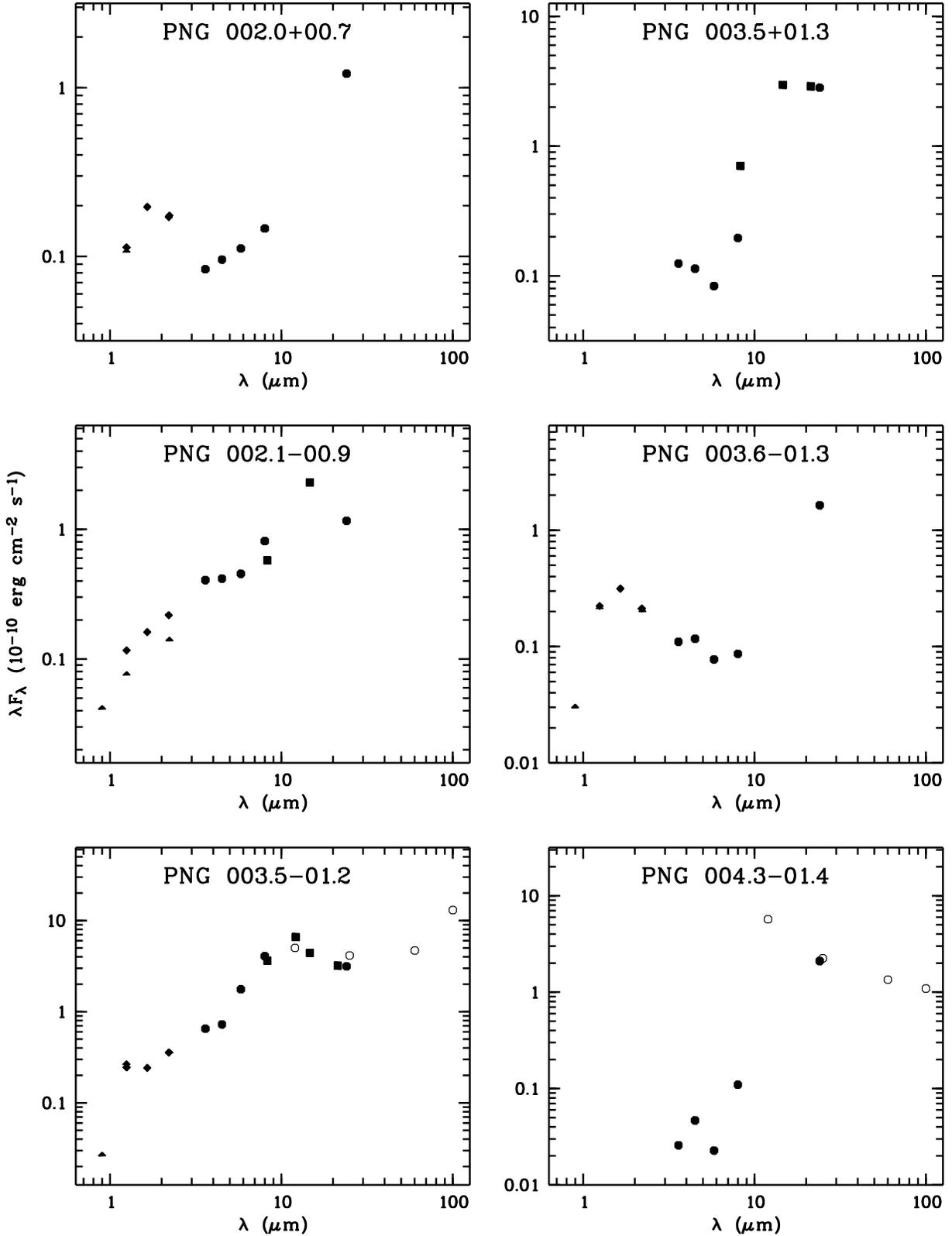,
height=22cm, bbllx=38, bblly=100, bburx=517, bbury=745, clip=, angle=0}
\caption{The SED of the PNs in the GLIMPSE~II sample. The filled triangles,
filled diamonds, filled circles, open circles, and filled squares
are from  the DENIS, 2MASS, GLIMPSE/MIPSGAL, {\it IRAS}, and {\it MSX} survey, 
respectively.  In the case of PNG $359.3-00.9$ (Hb 5), the ISO spectrum  is also plotted.
}
\label{sed}
\end{figure*}

\begin{figure*}
\centering
\epsfig{file=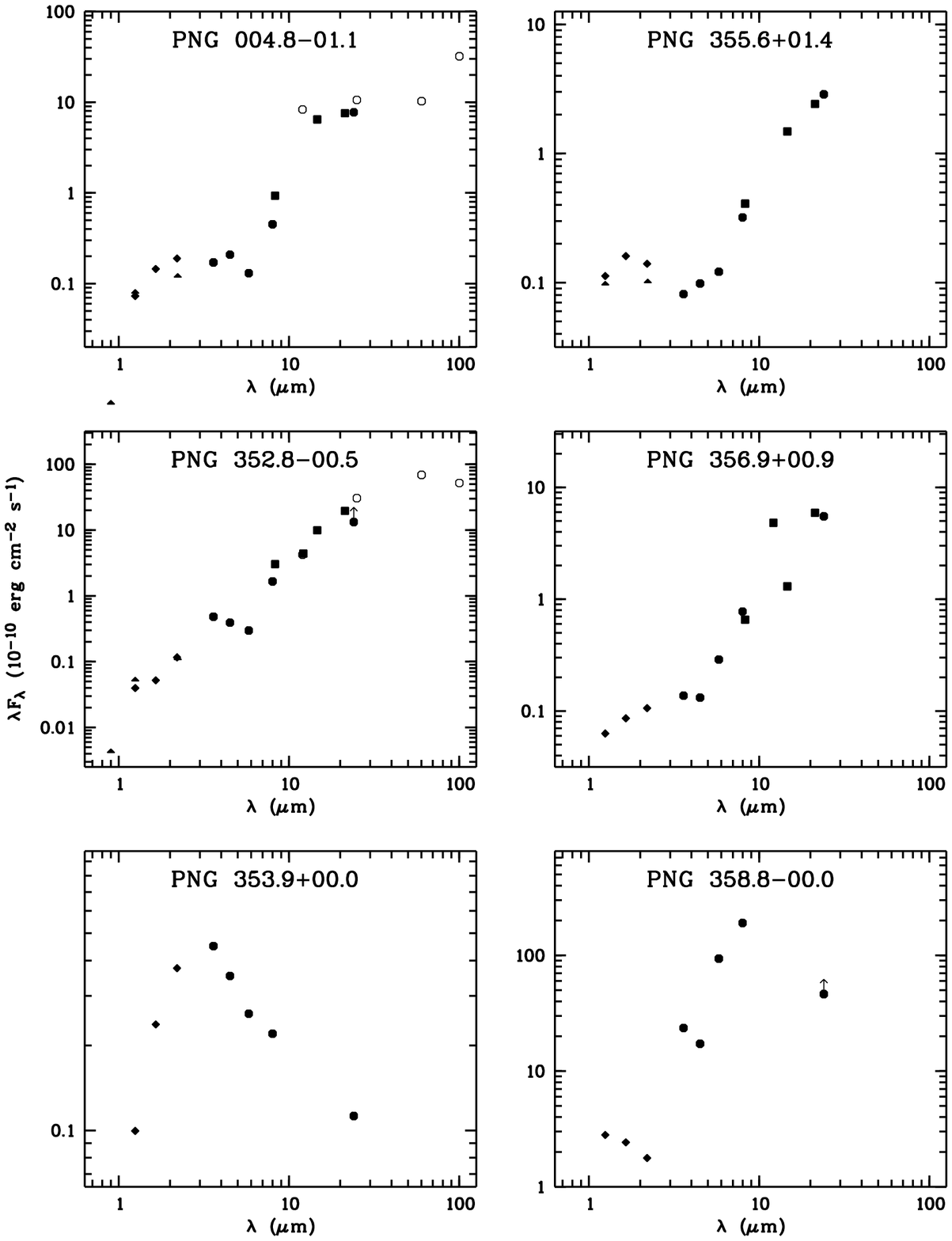,
height=22cm, bbllx=38, bblly=100, bburx=517, bbury=745, clip=, angle=0}
\end{figure*}

\begin{figure*}
\centering
\epsfig{file=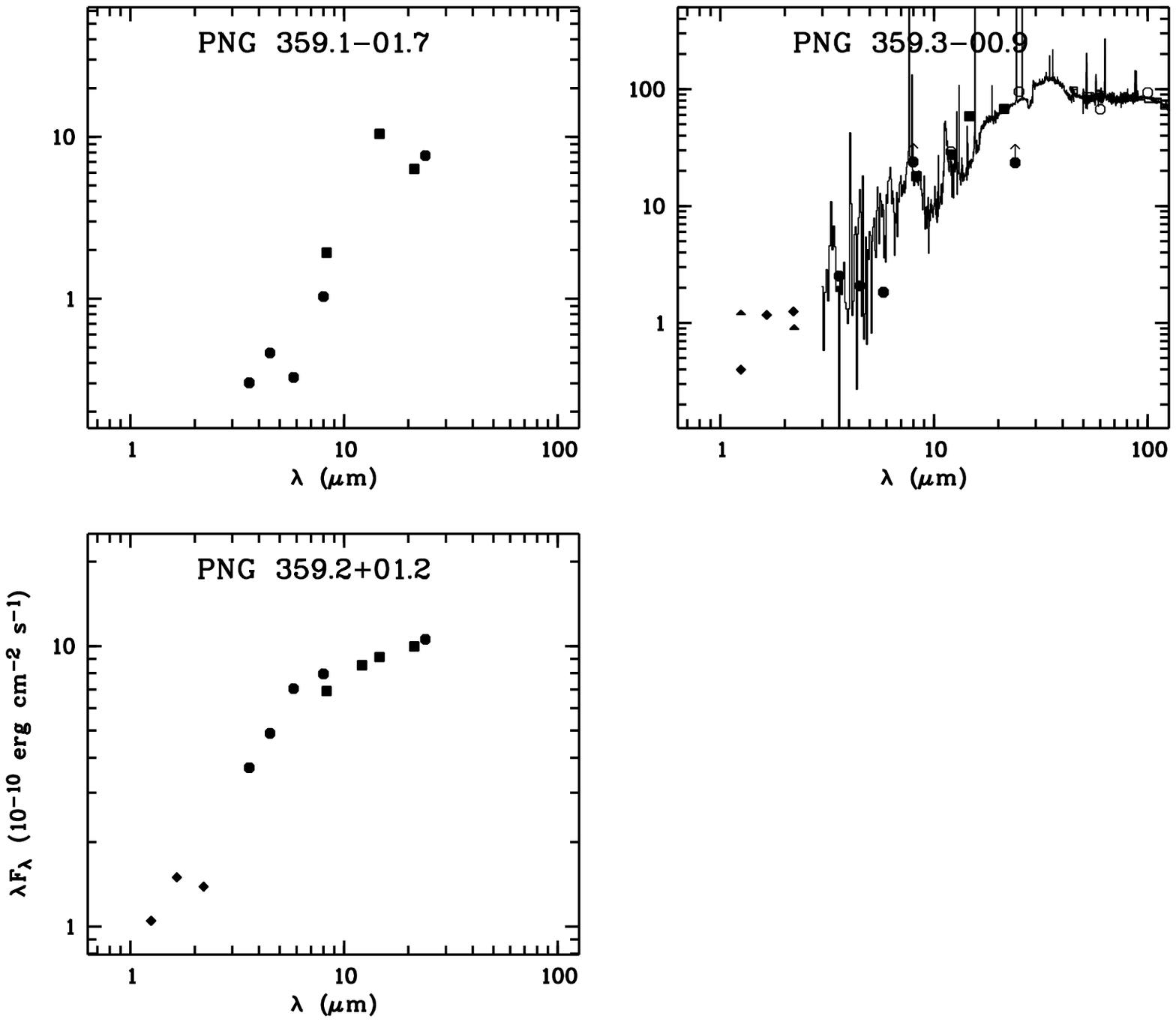,
height=22cm, bbllx=38, bblly=100, bburx=517, bbury=745, clip=, angle=0}
\end{figure*}

\begin{figure}
\epsfig{file=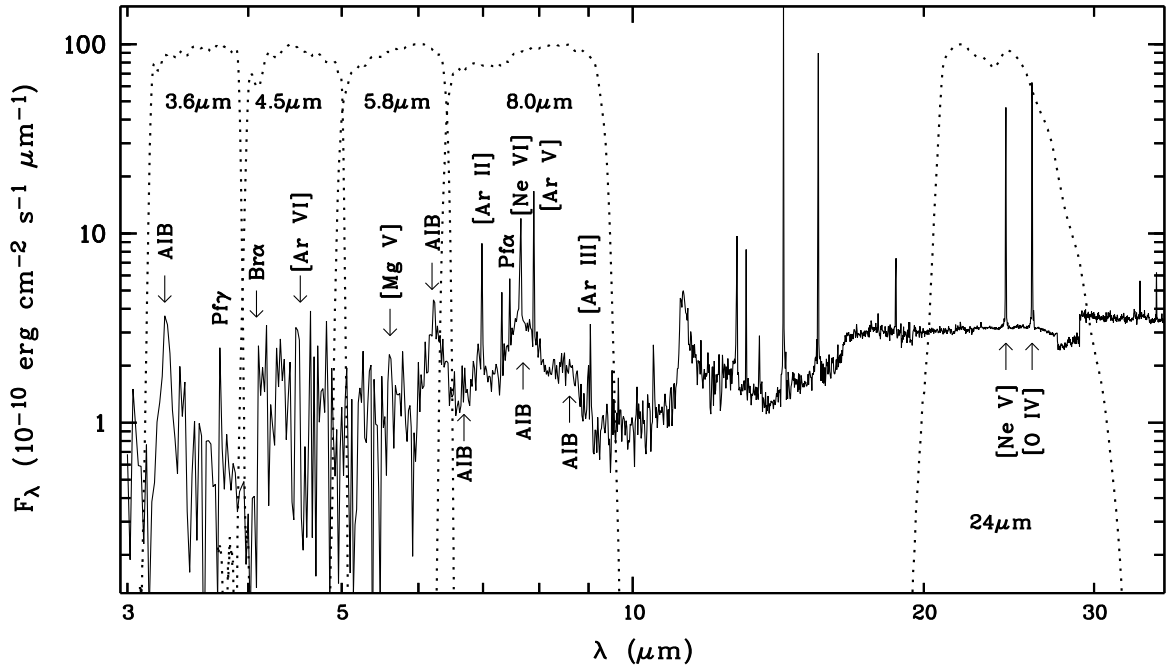, height=9cm,
bbllx=35, bblly=279, bburx=542, bbury=573, clip=, angle=0}
\caption{The {\it ISO} spectrum of Hb~5. The AIB emission and
strong atomic lines are identified. The normalized relative
spectral response curves for the IRAC and MIPS 24\,$\mu$m bands
are overlaid.
\protect\label{resp}}
\end{figure}

\begin{figure*}
\epsfig{file=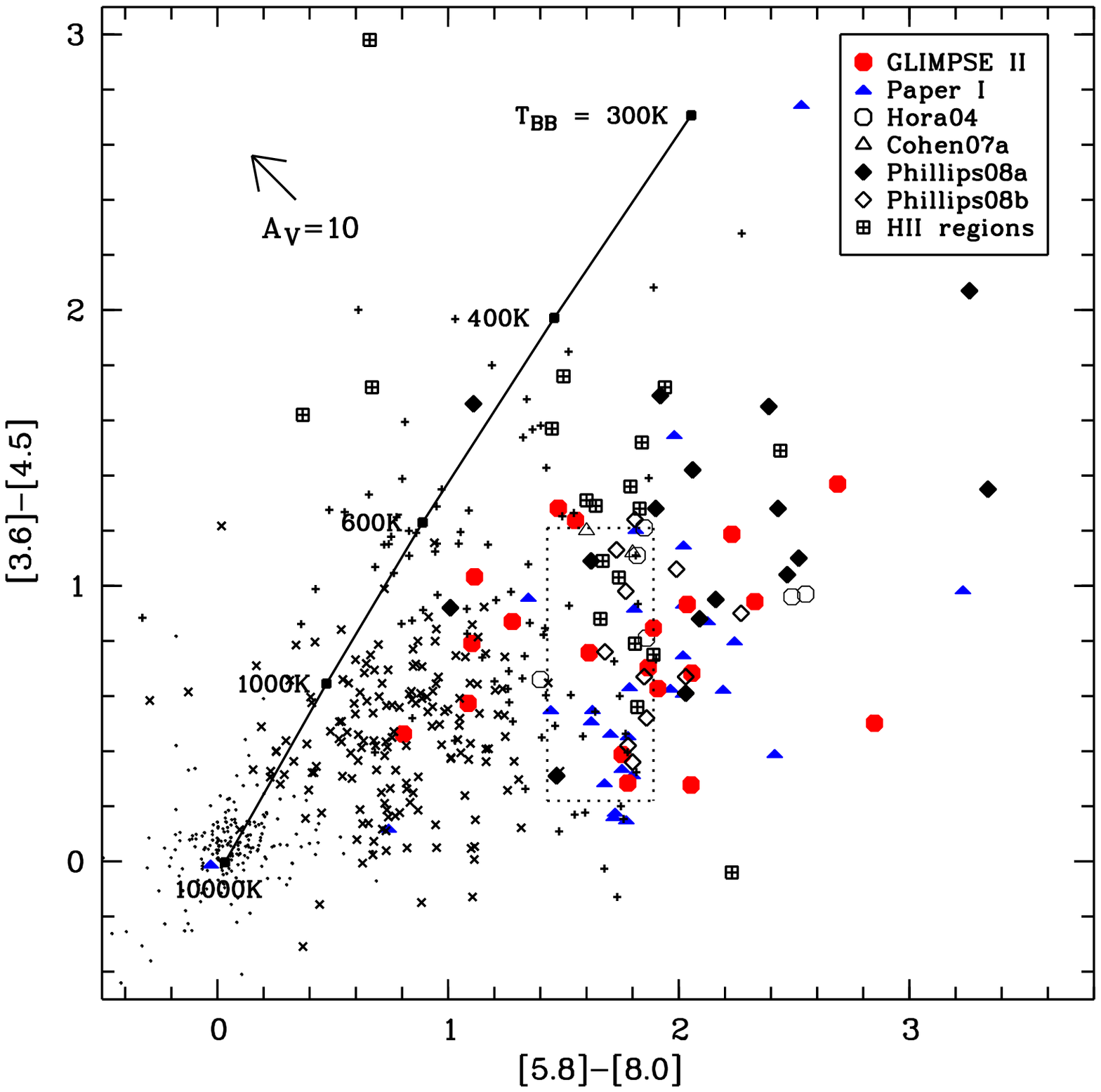, height=15cm,
bbllx=40, bblly=58, bburx=527, bbury=533, clip=, angle=0}
\caption{IRAC color-color plot ($[3.6]-[4.5]$ vs. $[5.8]-[8.0]$)
for the GLIMPSE~I and GLIMPSE~II PNs.  The observation results by 
\citet{hora04} and 
\citet{phillips08a,phillips08b} and
the theoretical calculations by \citet{cohen07a} are overlaid.
The enclosed zone by dotted lines represents that occupied by the PNs
studied by \citet{cohen07b}. We also show the colors of
UC H~II regions \citep{fuente08} and three types of YSOs 
\citep[star with infalling envelope---pluses, star with a disk---crosses, 
and post T-tauri star---points;][]{chavarria08}.  The filled squares connected
by a solid line denote the
prediction of blackbody radiation with various temperatures
$T_{\rm BB}$. The arrow in the upper left corner denotes a  
reddening vector of ${\rm A_v}=10$.
\protect\label{color}}
\end{figure*}

\begin{figure*}
\epsfig{file=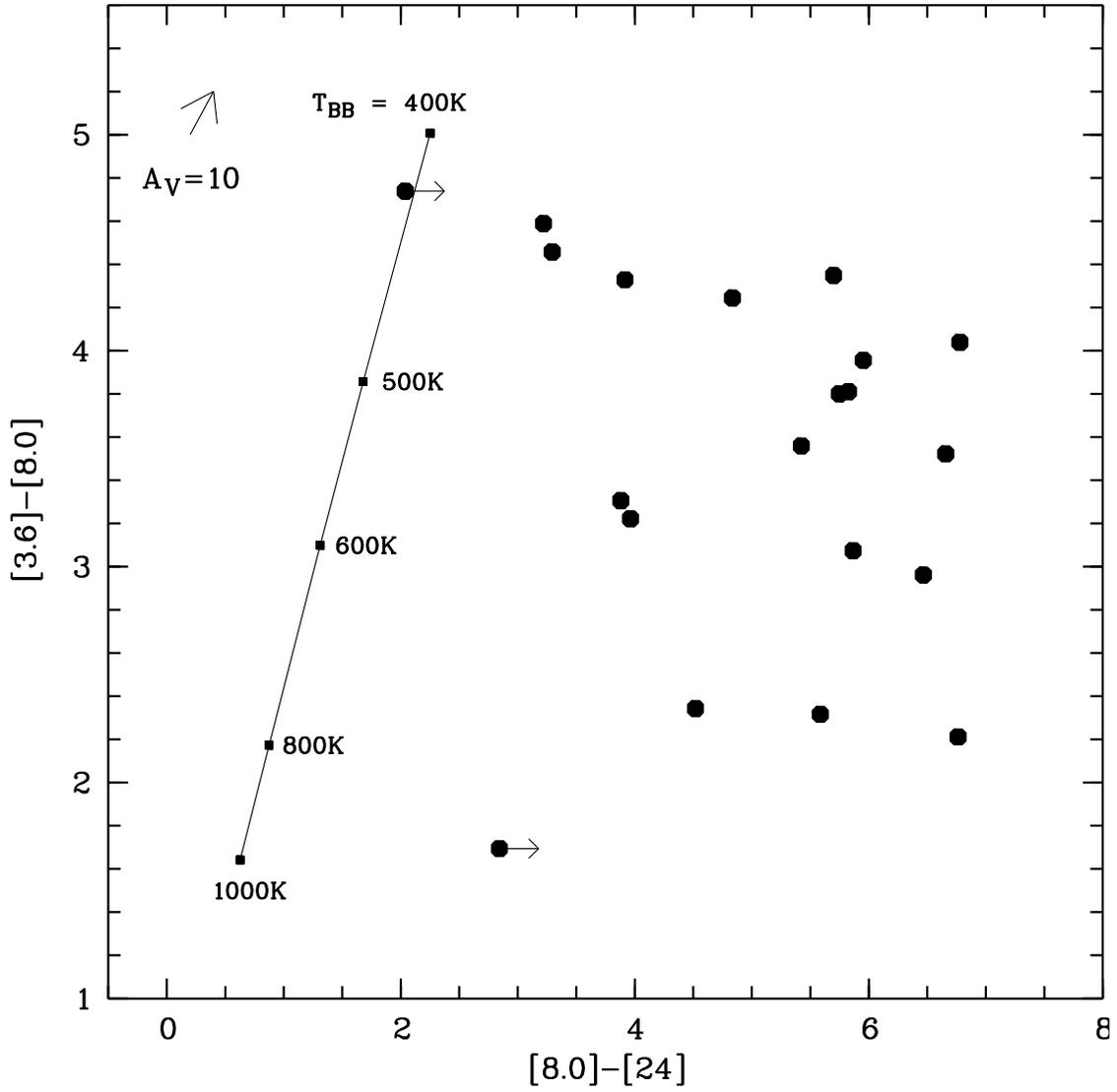, height=15cm,
bbllx=40, bblly=58, bburx=527, bbury=533, clip=, angle=0}
\caption{The $[3.6]-[8.0]$ versus $[8.0]-[24]$ color-color
diagram for the GLIMPSE~II PNs. Symbols are otherwise 
same as in Figure.~\ref{color}.
\protect\label{color2}}
\end{figure*}

\clearpage

\begin{figure*}
\centering
\epsfig{file=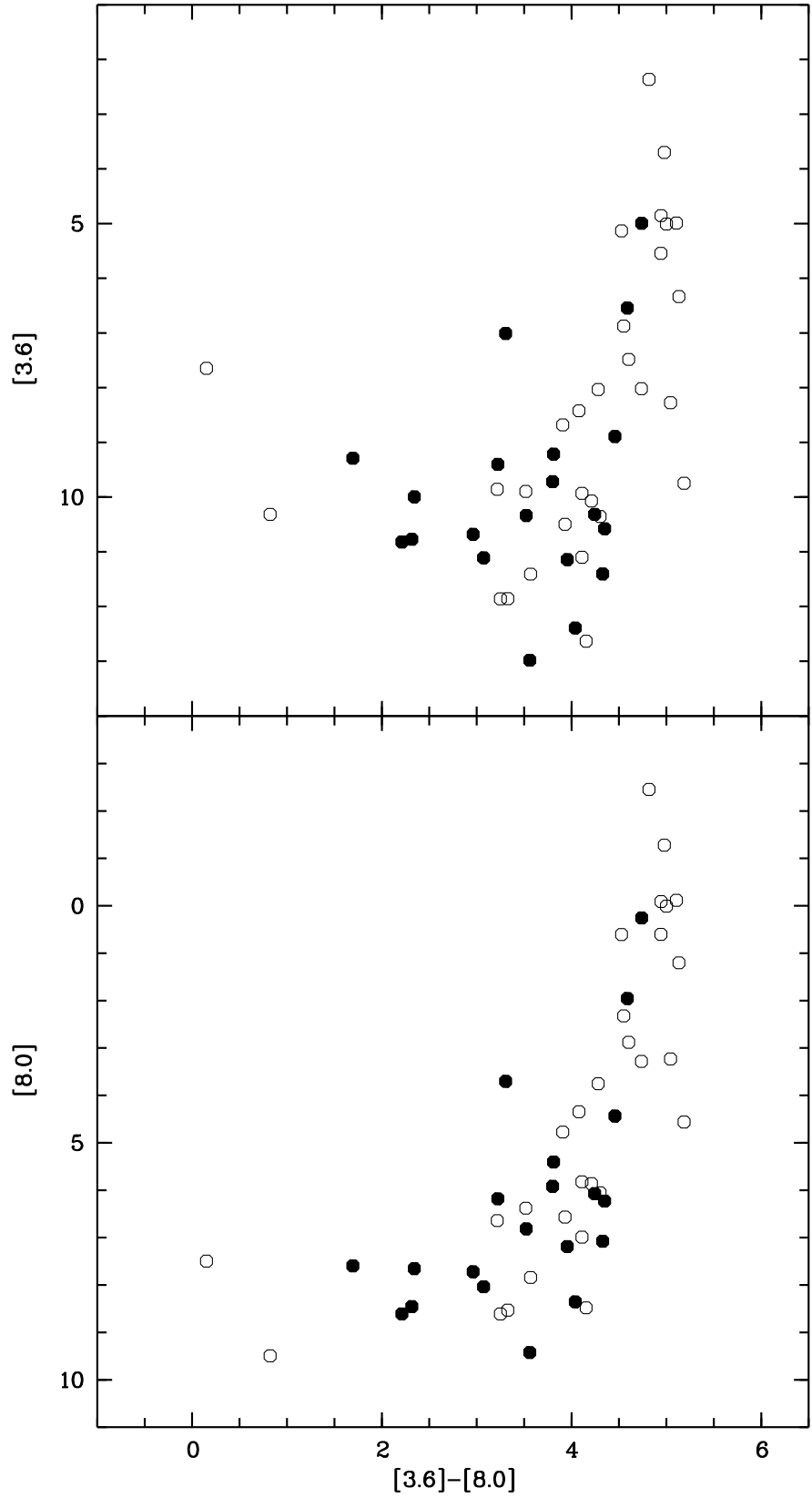, height=22cm,
bbllx=65, bblly=50, bburx=458, bbury=770, clip=, angle=0}
\caption{ The [3.6] versus $[3.6]-[8.0]$ (upper panel) 
and  [8.0] versus $[3.6]-[8.0]$  (lower panel) color-magnitude diagrams
for the GLIMPSE~I (open circles) and GLIMPSE~II (filled circles) PNs.
\protect\label{colmag}}
\end{figure*}

\clearpage

\begin{figure*}
\centering
\epsfig{file=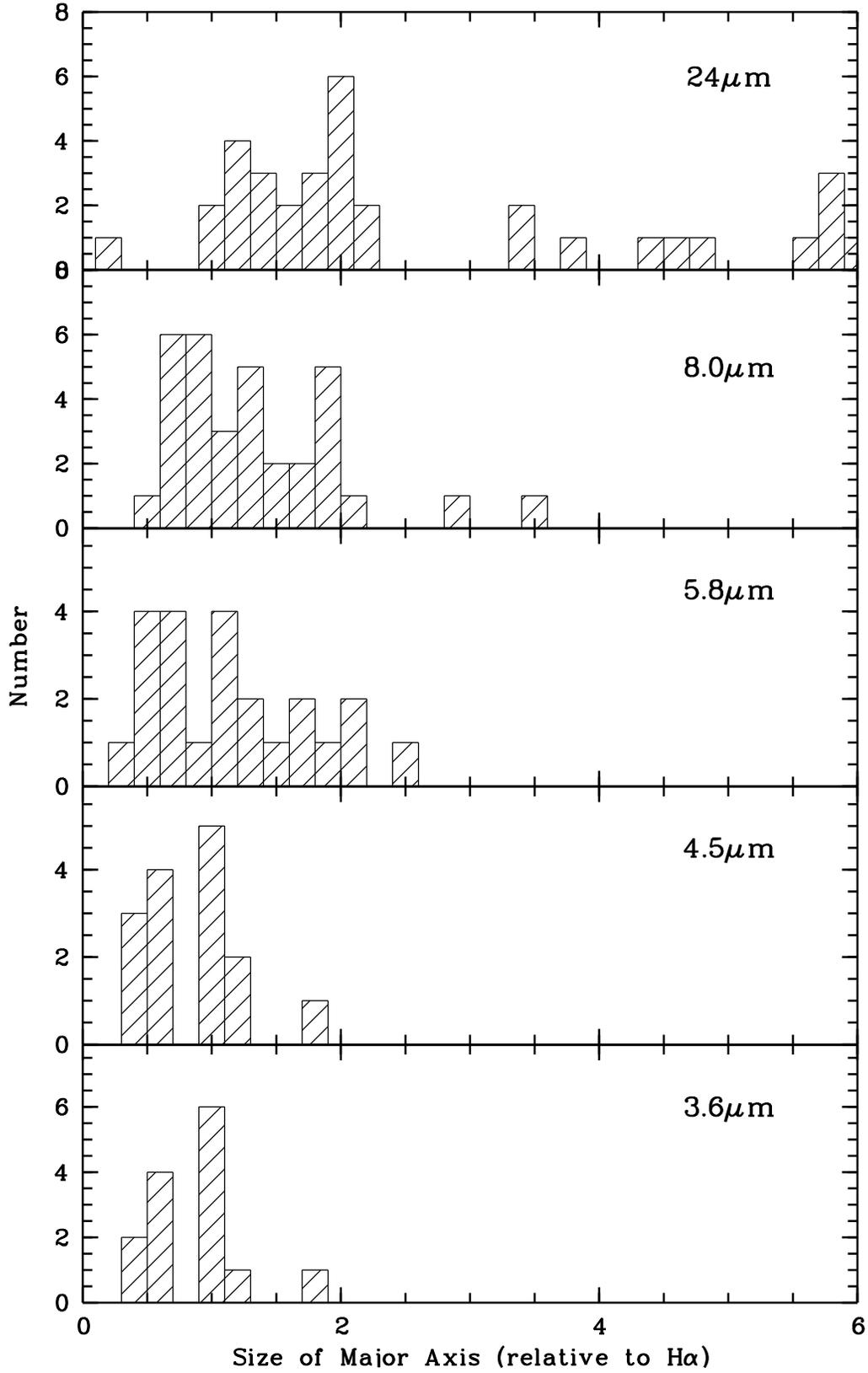, height=22cm,
bbllx=106, bblly=20, bburx=489, bbury=632, clip=, angle=0}
\caption{The histogram of the relative sizes of nebular major axises
measured in different wavelengths.
\protect\label{size}}
\end{figure*}

\clearpage

\begin{figure*}
\centering
\epsfig{file=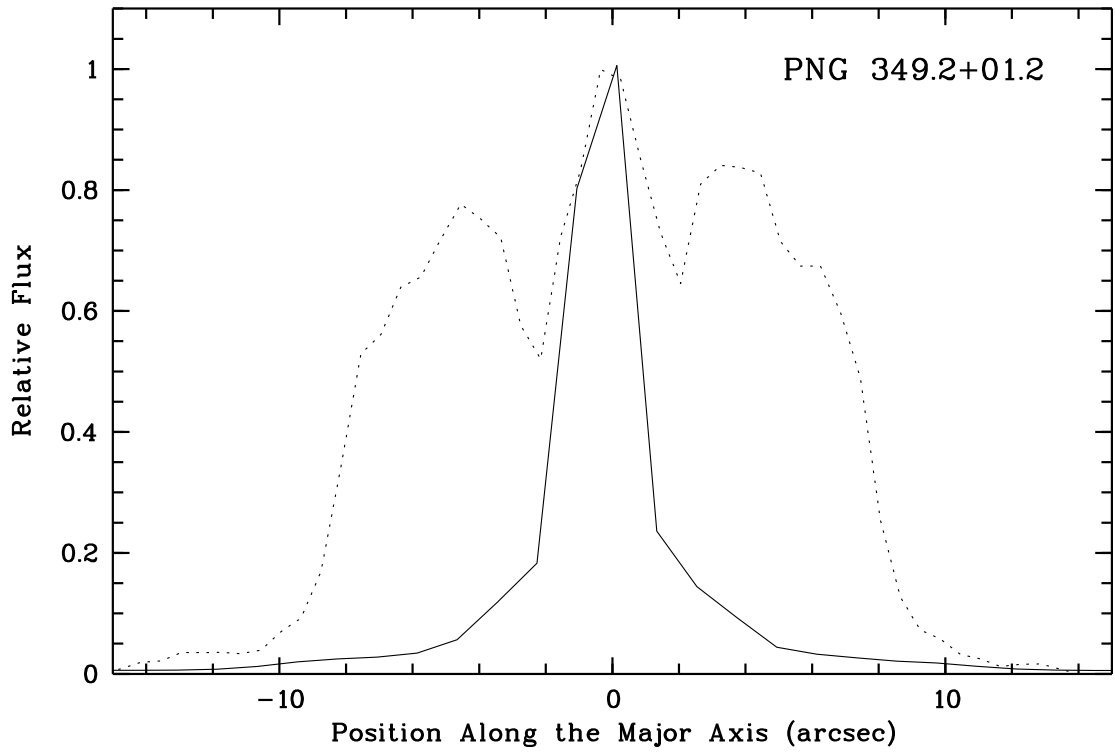, height=10cm,
bbllx=37, bblly=281, bburx=512, bbury=601, clip=, angle=0}
\caption{The flux distribution along the major axis of the
bipolar nebula PNG $359.2+01.2$ at H$\alpha$ (dotted line)
and 8.0\,$\mu$m (solid line) bands.
\protect\label{profile}}
\end{figure*}

\begin{figure*}
\center
\epsfig{file=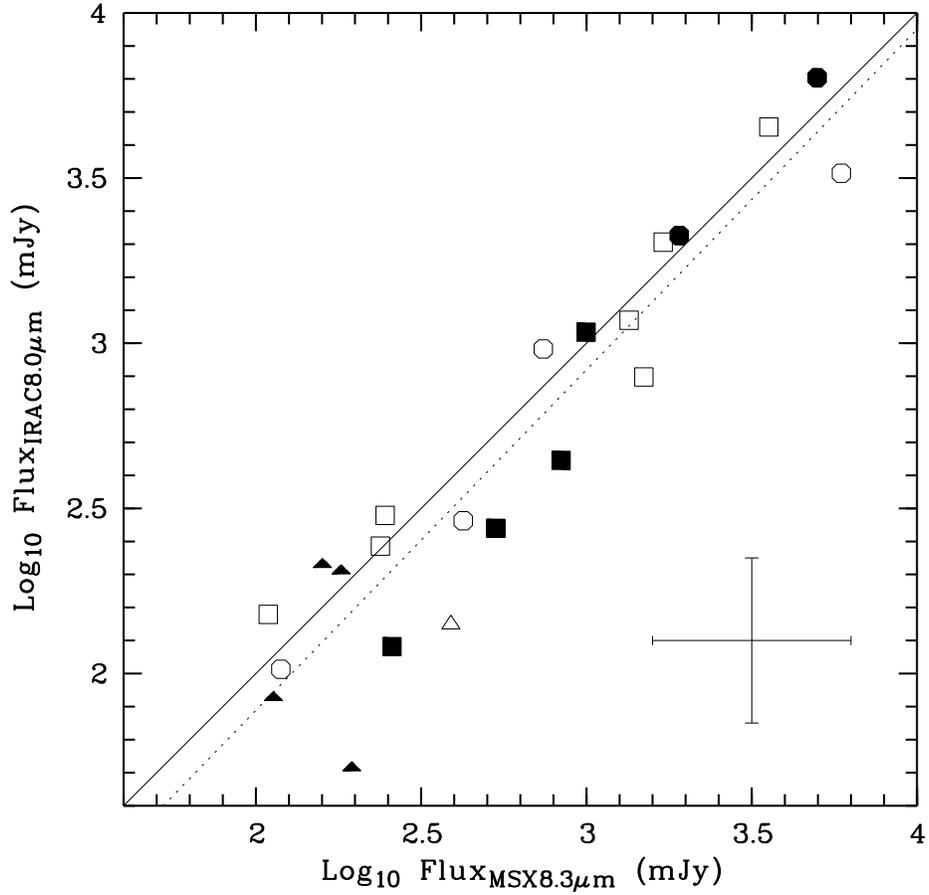, height=12cm,
bbllx=98, bblly=268, bburx=529, bbury=687, clip=, angle=0}
\caption{ IRAC 8.0\,$\mu$m versus MSX 8.3\,$\mu$m integrated fluxes for 
the GLIMPSE~I (open symbols; from Paper~I) and GLIMPSE~II (filled symbols) PNs. 
The triangles, squares, and circles denote the PNs with diameters ($R$)
in the ranges of $R\le10\arcsec$,  $10\arcsec < R\le20\arcsec$,
and $20\arcsec < R\le60\arcsec$, respectively.
The solid diagonal line is a $y=x$ plot. The dotted 
line represents a linear least-squares fit. The error bar is given
in the lower right corner.
\protect\label{compar1}}
\end{figure*}

\begin{figure*}
\center
\epsfig{file=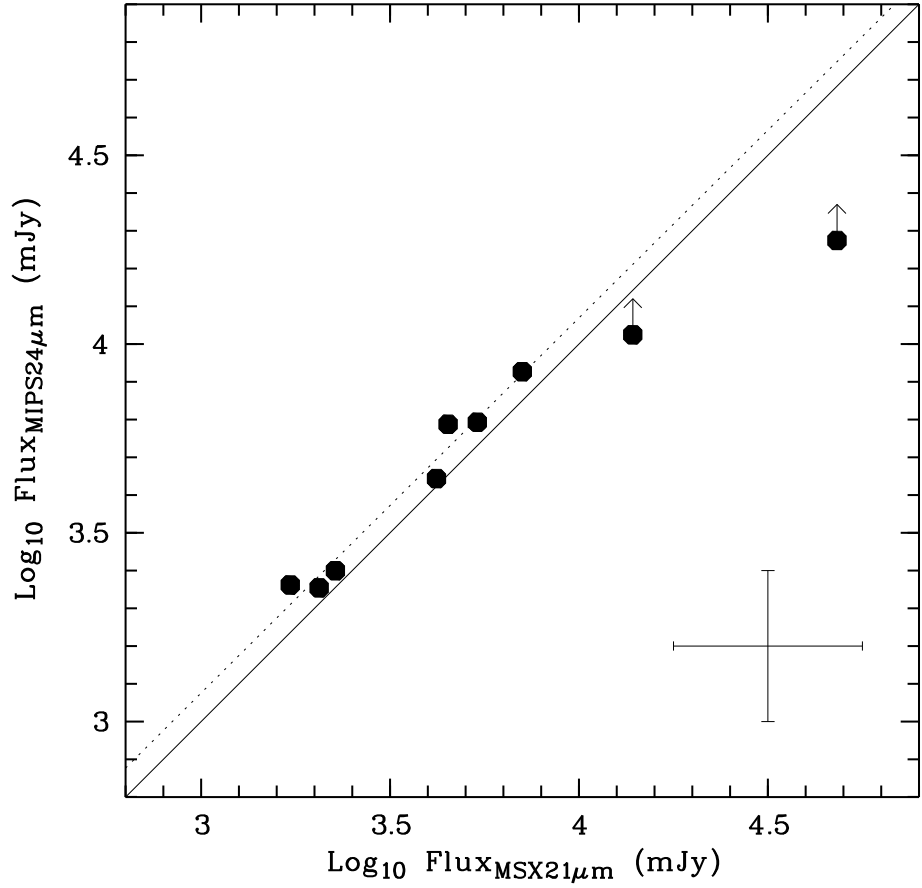, height=12cm,
bbllx=98, bblly=268, bburx=529, bbury=687, clip=, angle=0}
\caption{ IRAC 24\,$\mu$m 
versus MSX 21\,$\mu$m integrated fluxes for the GLIMPSE~II PNs.
The solid diagonal line is a $y=x$ plot. The dotted line
represents a linear least-squares fit. The error bar is given
in the lower right corner.
\protect\label{compar2}}
\end{figure*}

\begin{figure*}
\center
\epsfig{file=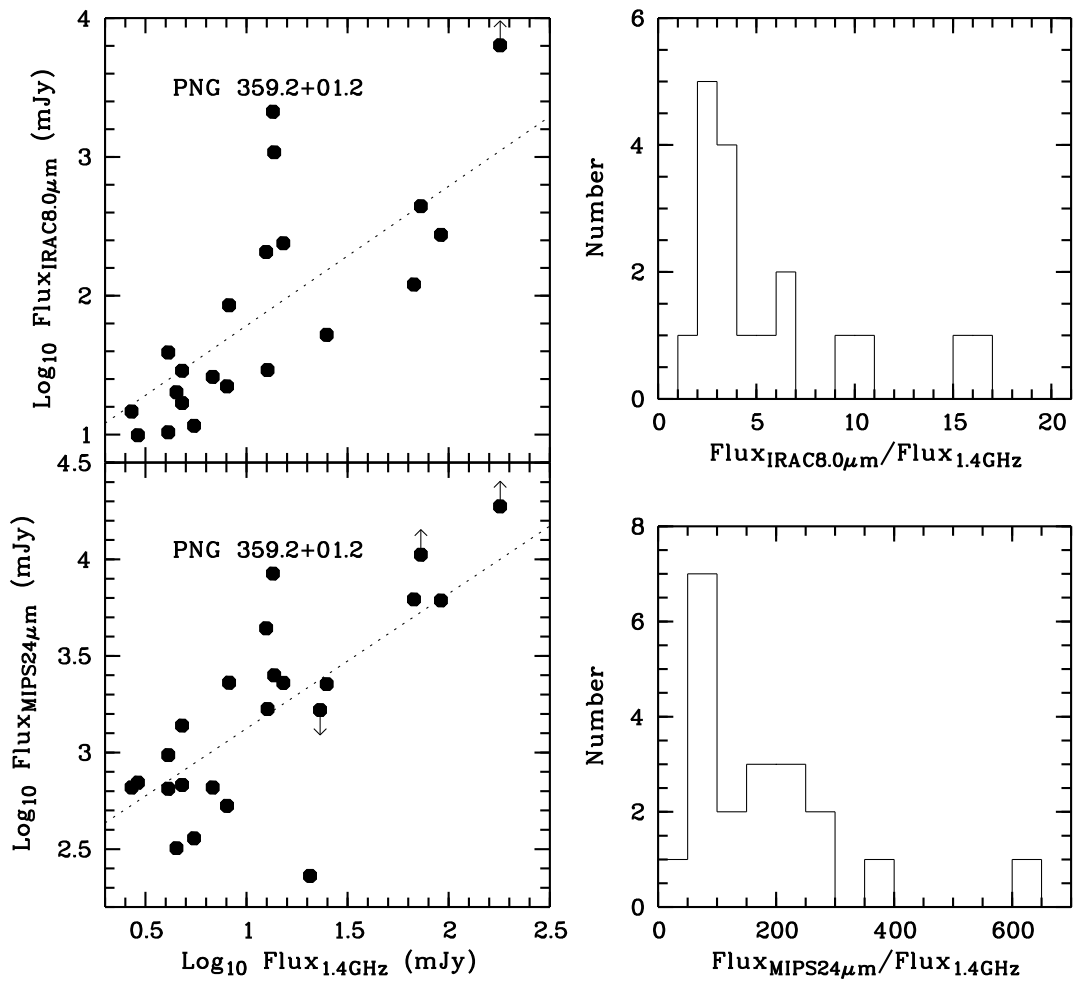, height=13cm,
bbllx=66, bblly=53, bburx=547, bbury=489, clip=, angle=0}
\caption{ {\it Left panels}: IRAC 8.0\,$\mu$m (upper panel) and MIPS 24\,$\mu$m
(lower panel) versus NVSS 1.4\,GHz integrated fluxes for the GLIMPSE~II PNs.
The dotted lines represent a linear fitting.
{\it Right panels}: the number distributions of the 
8.0$\mu$m/1.4GHz (upper panel; note that the two objects with
extremely large flux ratios are out of the range of the abscissa) and 
24$\mu$m/1.4GHz (lower panel) flux ratios.
\protect\label{radio}}
\end{figure*}

\newpage

\clearpage

\begin{deluxetable}{cl}
\tablecaption{Known PNs.
\label{name}}
\tablewidth{0pt}
\tablehead{
\colhead{Object} & \colhead{Other Name}\\
}
\startdata
PNG $000.5+01.9$ & PN~K~6-7\\
PNG $001.0-01.9$ & PN~K~6-35\\
PNG $001.0+01.9$ & Hen~2-264, PN~K1-4, PN~Sa~3-69\\
PNG $002.2+00.5$ & Terz~N~2337  \\
PNG $358.2-01.1$ & PN~Al~2-L, PN~Bl~1-D  \\
PNG $358.8-00.0$ & Terz~N~2022  \\
PNG $359.1-01.7$ & Hen~1-191, PN~M~1-29, PN~Sa~2-250  \\
PNG $359.2+01.2$ & PN~PM~1-166, 19W32  \\
PNG $359.3-00.9$ & Hen~2-286,  PN~Bl~1-E, PN~Hb~5, PN~Sa~2-244  \\
\enddata
\end{deluxetable}

\pagestyle{empty}

\textheight 24cm

\begin{deluxetable}{lcc@{\extracolsep{0.1in}}rrrrr@{\extracolsep{0.1in}}ccccccc}
\rotate
\tablecaption{{\it Spitzer} observations of PNs in GLIMPSE~II.
\label{flux}}
\tabletypesize{\scriptsize}
\tablewidth{0pt}
\tablehead{ 
      & \multicolumn{2}{c}{Obs. Coord. (J2000)}&\multicolumn{5}{c}{Flux (mJy)} & \multicolumn{4}{c}{Size (arcsec)} \\
\cline{2-3} \cline{4-8}\cline{9-14}
\colhead{Object} & \colhead{R.A}& \colhead{Decl.}&\colhead{3.6$\mu$m}& \colhead{4.5$\mu$m} & \colhead{5.8$\mu$m} & \colhead{8.0$\mu$m} & \colhead{24$\mu$m} &
\colhead{3.6$\mu$m}& \colhead{4.5$\mu$m} & \colhead{5.8$\mu$m} & \colhead{8.0$\mu$m} & \colhead{24$\mu$m} & \colhead{H$\alpha^a$} & \colhead{$\alpha_{\rm IRAC}$}\\ 
}
\startdata
PNG 000.0$-01.3$&      17 51 00.4 &$-29$ 33 51&...  & ...  &  9.1   &  20.2  & 320      & ...         & ...         & $9\times$11 & $10\times$13 & 20           & 10.0$\times$12.0 &...\\
PNG 000.1$-01.7$&      17 52 49.0 &$-29$ 41 56&...  & ...  &  ...   &   9.9  & 699      & ...         & ...         & ...         &           12 & 20           & 11.0$\times$17.0 &...\\
PNG 000.3$-01.6$&      17 52 52.1 &$-29$ 30 04&...  & ...  &  ...   &  10.4  & 650      & ...         & ...         & ...         &           8  & 13           &             5.6  &...\\
PNG 000.5$+01.9$&      17 39 31.4 &$-27$ 27 47&...  & ...  &  ...   &  26.1  & 660      & ...         & ...         & ...         &  $9\times$11 & 15           &  $6.0\times$9.5  &...\\
PNG 000.9$-01.0$&      17 51 43.3 &$-28$ 38 59&...  & ...  &  ...   &  23.1  & 650      & ...         & ...         & ...         &           9  & 15           &             6.0  &...\\
PNG 000.9$+01.8$&      17 40 51.7 &$-27$ 08 48&...  & ...  &  ...   &  12.0  & 153      & ...         & ...         & ...         &  $4\times$6  & 11           &  5.5$\times$7.0  &...\\
PNG 001.0$-01.9$&      17 55 43.9 &$-29$ 04 05&...  & ...  &  ...   &  16.9  & 1380     & ...         & ...         & ...         & $11\times$15 & 35           & $15.0\times$17.0 &...\\
PNG 001.0$+01.9$&      17 40 26.8 &$-27$ 01 09&...  & ...  &  ...   &  ...   & 230      & ...         & ...         & ...         &  ...         & 13           & $53.5\times$78.2 &...\\
PNG 001.6$-01.1$&      17 54 13.4 &$-28$ 05 17&...  & ...  &  ...   &  30.3  & 524      &   ...       &         ... &         ... &  $9\times$10 & 27           & $11.0\times$15.0 &...\\
PNG 002.0$+00.7$&      17 47 28.3 &$-26$ 49 48&10.1 & 14.4 & 21.6   &  39.1  & 970      &   4         &          4  &          7  &           8  & 25           &              4.0 & 0.69\\
PNG 002.1$-00.9$&      17 54 18.8 &$-27$ 36 36&48.7 & 62.6 & 87.9   & 216.2  & 930      &   6         &          6  &          9  &           12 & 28           &              6.0 & 0.85\\
PNG 002.1$-01.1$&      17 55 10.4 &$-27$ 41 40&...  & ...  &  7.3   &  14.7  & 660      & ...         & ...         &          4  &           6  & 31           &              8.0 &...\\
PNG 002.2$+00.5$&      17 48 45.6 &$-26$ 43 31&21.0 & 42.0 &102.6   & 239.0  & 2295     & ...         & ...         &$23\times$32 & $26\times$35 & 32           & 17.0$\times$28.0 & 2.08\\
PNG 002.2$-01.2$&      17 55 45.6 &$-27$ 39 41&...  &  5.4 &  5.8   &  22.3  & 530      & ...         & $4\times$5  & $5\times$7  & $10\times$12 & 32           & 11.0$\times$14.5 &...\\
PNG 003.4$+01.4$&      17 48 15.6 &$-25$ 15 14&...  & ...  & ...    &  19.4  & 67       & ...         & ...         & ...         &           10 & 14           & 12.5$\times$14.0 &...\\
PNG 003.5$-01.2$&      17 58 37.0 &$-26$ 28 47&78.0 &108.8 &339.8   &1080.1  & 2510     &   6         &          6  &          11 &           11 & 30           &  4.0$\times$5.0  &2.44\\
PNG 003.5$+01.3$&      17 48 41.7 &$-25$ 11 34&15.0 & 17.1 & 16.2   &  52.4  & 2260     &   5         &          5  &          9  &           10 & 30           &  4.0$\times$5.0  &0.47\\
PNG 003.6$-01.3$&      17 59 11.0 &$-26$ 30 23&13.2 & 17.5 & 15.0   &  23.1  & 1310     &   6         &          7  &          8  &           10 & 30           &             14.0 &$-0.41$\\
PNG 003.7$+00.5^b$&    17 52 08.9 &$-25$ 27 43& ... & ...  & ...    &  ...   &$<1660$   & ...         & ...         & ...         & ...          & ...          &             10.2 &...\\
PNG 004.3$-01.4^c$&    18 01 18.9 &$-25$ 53 21& 3.1 & 7.0  &  4.4   &  29.2  & 1680     &   4         &          4  &          5  &            8 & 34           &  5.0$\times$6.0  &1.44\\
PNG 004.8$-01.1$&      18 01 17.0 &$-25$ 22 37& 20.6& 31.4 & 25.3   & 120.6  & 6205     & $5\times$8  & $5\times$8  & $7\times$10 & $10\times$13 & 35           &  6.0$\times$8.0  &1.00\\
PNG 006.1$+00.8$&      17 56 33.2 &$-23$ 11 47& 13.8& 11.4 &  7.2   &  26.6  & 510      &  ...        & ...         & ...         &           11 & 32           &             9.0  &$-0.32$\\
PNG 352.1$-00.0^{d,e}$&17 25 33.4 &$-35$ 36 25&  678& 1412 & 4873   & 10599  &23000     &  ...        & ...         &$73\times$96 &$107\times$156& 40$\times$50 &             45.0 &2.55\\
PNG 352.8$-00.5^f$&    17 29 37.5 &$-35$ 13 46& 57.9& 58.8 & 57.5   & 441.9  &$>10580$  & $5\times$7  & $7\times$10 &$13\times$17 & $13\times$17 & 34           &  9.0$\times$18.0 &1.43\\
PNG 353.9$+00.0^e$&    17 30 29.5 &$-34$ 00 37& 54.0& 52.9 & 50.1   &  58.7  & 90       &   5         &          6  &          7  &           10 & 11           &  4.0$\times$5.0  &$-0.91$\\
PNG 355.6$-01.4$&      17 40 55.2 &$-33$ 24 18&  ...&   ...&  ...   &  11.6  & 360      &  ...        & ...         & ...         &  $6\times$9  & 20           &             9.4  &...\\
PNG 355.6$+01.4$&      17 29 11.0 &$-31$ 52 45&  9.8& 14.8 & 23.5   &  85.5  & 2300     &   4         &          4  &          7  &           10 & 30           &  5.0$\times$6.0  &1.67\\
PNG 356.0$-01.4$&      17 41 33.4 &$-33$ 02 15&  ...&   ...&  7.2   &  28.9  & 680      &  ...        & ...         &          5  &           11 & 28           &  7.0$\times$8.0  &...\\
PNG 356.9$+00.9$&      17 34 34.3 &$-31$ 02 08& 16.5& 19.8 & 55.8   & 206.8  & 4400     &   5         &          5  &          10 &           10 & 30           &             5.0  &2.31\\
PNG 357.4$-01.3^{e,g}$&17 44 56.2 &$-31$ 49 19&  ...&   ...&  ...   &  ...   & ...      &  ...        & ...         &          300&           300& 300          &             256  &...\\
PNG 357.5$+01.3$&      17 34 26.0 &$-30$ 15 18&  7.7&  6.4 & 33.0   &  94.7  & 390      &   4         &          4  &          8  &           10 & 15           &  4.0$\times$7.0  &2.51\\
PNG 357.7$+01.4$&      17 34 46.6 &$-30$ 04 21&  1.8&  2.2 &  3.5   &  10.9  & 180      &  ...        & ...         &          4  &           6  & 15           &  6.0$\times$10.0 &1.27\\
PNG 358.2$-01.1$&      17 46 03.1 &$-31$ 03 36& 28.2& 30.6 & 36.7   &  55.7  & 400      &  ...        & ...         &          17 &           19 & 25           & 18.8$\times$22.6 &$-0.14$\\
PNG 358.8$-00.0^{e,f}$&17 42 42.6 &$-29$ 51 35& 2823& 2582 &18057   & 50662  &$>37000$  &   50        &          50 &          70 &           80 & 55           & 11.3$\times$28.3 &3.04\\
PNG 359.1$-01.7$&      17 50 18.0 &$-30$ 34 55& 36.4&  69.5& 63.3   &  275.3 & 6130     &   9         &          9  &          13 &           15 & 35           & 17.0$\times$18.8 &1.31\\
PNG 359.2$+01.2$&      17 39 02.9 &$-28$ 56 37&442.1& 732.0&1361.6  & 2119.6 & 8450     & $8\times$16 & $8\times$16 & $9\times$17 & $13\times$22 & 30$\times$42 &  9.5$\times$26.5 &0.99\\
PNG 359.3$-00.9^{f,h}$&17 47 56.1 &$-29$ 59 41&303.1& 312.6& 354.8  &$>6369$ &$>18800$  &   12        &          12 &          23 &           23 & 30           & 28.3$\times$60.3 &...\\
\enddata
\tablenotetext{a}{Taken from \citet{parker01,parker06} and \citet{miszalski08}.}
\tablenotetext{b}{Badly blended with a bright star, which saturates all the IRAC bands.}
\tablenotetext{c}{The MIPS image is probably blended with two 
unresolved field stars.}
\tablenotetext{d}{The {\it Spitzer} images reveal diffuse tail.  
The flux and size of the central bright region are given for the MIPS24 band. }
\tablenotetext{e}{Non-PNs.}
\tablenotetext{f}{The MIPS images are saturated.}
\tablenotetext{g}{It shows large circle structure with ambiguous edge, and the given sizes are rough.}
\tablenotetext{h}{The 8.0\,$\mu$m image is saturated.}
\end{deluxetable}

\clearpage
\begin{deluxetable}{cccc@{\extracolsep{0.1in}}ccc@{\extracolsep{0.1in}} 
cccc@{\extracolsep{0.1in}}cccc} 
\rotate 
\tablecaption{Other flux measurements \label{other}} 
\tabletypesize{\scriptsize} 
\tablewidth{9.6in} 
\tablehead{ 
 & 
\multicolumn{3}{c}{DENIS} & 
\multicolumn{3}{c}{2MASS} & 
\multicolumn{4}{c}{{\it MSX}} & 
\multicolumn{4}{c}{{\it IRAS}} \\ 
\cline{2-4}\cline{5-7}\cline{8-11}\cline{12-15} 
& 
\multicolumn{1}{c}{I} & 
\multicolumn{1}{c}{J} & 
\multicolumn{1}{c}{K} & 
\multicolumn{1}{c}{J} & 
\multicolumn{1}{c}{H} & 
\multicolumn{1}{c}{K} & 
\multicolumn{1}{c}{8.28\,$\mu$m} & 
\multicolumn{1}{c}{12.13\,$\mu$m} & 
\multicolumn{1}{c}{14.65\,$\mu$m} & 
\multicolumn{1}{c}{21.3\,$\mu$m} & 
\multicolumn{1}{c}{12\,$\mu$m} & 
\multicolumn{1}{c}{25\,$\mu$m} & 
\multicolumn{1}{c}{60\,$\mu$m} & 
\multicolumn{1}{c}{100\,$\mu$m} \\ 
\multicolumn{1}{c}{Object}& 
\multicolumn{1}{c}{(mag)} & 
\multicolumn{1}{c}{(mag)} &
\multicolumn{1}{c}{(mag)} &
\multicolumn{1}{c}{(mag)} & 
\multicolumn{1}{c}{(mag)} &
\multicolumn{1}{c}{(mag)} &
\multicolumn{1}{c}{(Jy)} & 
\multicolumn{1}{c}{(Jy)} &
\multicolumn{1}{c}{(Jy)} &
\multicolumn{1}{c}{(Jy)} &
\multicolumn{1}{c}{(Jy)} &
\multicolumn{1}{c}{(Jy)} &
\multicolumn{1}{c}{(Jy)} &
\multicolumn{1}{c}{(Jy)} 
} 
\startdata 
PNG 002.0$+00.7$&  ...    &  13.902  & 11.764  &  13.858  & 12.486  & 11.82    & ...    &  ...    & ...   & ...     &  ...    &  ...   &   ...   &    ...\\
PNG 002.1$-00.9$&  15.616 &  14.268  & 12.009  &  13.827  & 12.702  & 11.56    & 0.159  &  ...    & 1.121 & ...     &  ...    &  ...   &   ...   &    ...\\
PNG 003.5$-01.2$&  16.089 &  12.894  & 10.980  &  13.020  & 12.264  & 11.03    & 0.998  &  2.653  & 2.153 &  2.266  &  1.996  & 3.438  & 9.337   &43.230\\
PNG 003.5$+01.3$&  ...    &  ...     & ...     &  ...     & ...     & ...      & 0.195  &  ...    & 1.450 &  2.052  &  ...    &  ...   &   ...   &    ...\\
PNG 003.6$-01.3$&  15.963 &  13.137  & 11.606  &  13.123  & 11.978  & 11.59    & ...    &  ...    & ...   & ...     &  ...    &  ...   &   ...   &    ...\\
PNG 004.3$-01.4$&  ...    &  ...     & ...     &  ...     & ...     & ...      & ...    &  ...    & ...   & ...     &  2.272  & 1.864  & 2.691   & 3.623 \\
PNG 004.8$-01.1$&  17.951 &  14.203  & 12.166  &  14.333  & 12.815  & 11.71    & 0.258  &  ...    & 3.162 &  5.381  &  3.329  & 8.839  & 20.550  & 106.900\\
PNG 352.8$-00.5$&  18.071 &  14.654  & 12.250  &  14.995  & 13.929  & 12.24    & 0.838  & 1.774   & 4.84  & 13.878  &  1.693  & 25.430 & 137.400 & 172.900\\
PNG 353.9$+00.0$&  ...    &  ...     & ...     &  13.999  & 12.284  & 10.97    & ...    &  ...    & ...   & ...     &  ...    &  ...   &   ...   &    ...\\
PNG 355.6$+01.4$&  ...    &  14.002  & 12.361  &  13.865  & 12.708  & 12.04    & 0.113  & ...     & 0.725 & 1.720   &  ...    &  ...   &   ...   &    ...\\
PNG 356.9$+00.9$&  ...    &  ...     & ...     &  14.495  & 13.384  & 12.34    & 0.181  & 1.946   & 0.635 & 4.195   &  ...    &  ...   &   ...   &    ...\\
PNG 358.8$-00.0$&  ...    &  ...     & ...     &  10.375  & 9.763   & 9.285    & ...    &  ...    & ...   & ...     &  ...    &  ...   &   ...   &    ...\\
PNG 359.1$-01.7$&  ...    &  ...     & ...     &  ...     & ...     & ...      & 0.532  & ...     & 5.100 &   4.500 &  ...    &  ...   &   ...   &    ...\\
PNG 359.2$+01.2$&  ...    &  ...     & ...     &  11.442  & 10.283  & 9.550    & 1.908  & 3.452   & 4.459 &   7.076 &  ...    &  ...   &   ...   &    ...\\
PNG 359.3$-00.9$&  13.149 &  11.265  & 9.980   &  11.037  & 10.852  & 9.854    & 4.985  & 11.118  & 28.664&   48.165&  11.680 &  79.240&  134.500&  311.800\\
\enddata 
\end{deluxetable}

\end{document}